%% file: paper.tex
\documentclass[10pt]{sig-alternate-10pt}
\usepackage{times}

\include{psfig}  
\usepackage{siunitx}
\usepackage{epsfig}
\usepackage{algorithm }
\usepackage{xspace}
\usepackage{amssymb}
\usepackage{amsmath}
\usepackage{algpseudocode}
\usepackage{float}
\usepackage{subcaption}
\usepackage{graphicx}
\usepackage[update,prepend]{epstopdf}
\usepackage{flushend}
\usepackage{color}
\usepackage[hyphens,spaces,obeyspaces]{url}
\usepackage{array}
\usepackage{gensymb}
\usepackage{soul}

\floatstyle{plain}
\newfloat{deneme}{t!}

\newcommand{\ie}{\textit{i.e.}\xspace}

\newcommand{\system}{\textsc{SkyHaul}\xspace}

\algdef{SE}[DOWHILE]{Do}{doWhile}{\algorithmicdo}[1]{\algorithmicwhile\ #1}%

\begin{document}



\newfont{\titlefont}{phvb8t at 17pt}
\twocolumn[
\centerline{\titlefont SkyHaul: An Autonomous Gigabit Network Fabric In The Sky}
\vspace{30pt}
\centerline{Ramanujan K Sheshadri, Eugene Chai, Karthikeyan Sundaresan, Sampath Rangarajan}
\vspace{12pt}
\centerline{NEC Laboratories America, Inc.}
\vspace{12pt}
\centerline{(ram,eugene,karthiks,sampath)@nec-labs.com}
]
\medskip

\begin{abstract}
\input{abstract1.tex}
\end{abstract}

\section{Introduction}
\label{sec:intro}
\input{intro-v7.tex}

\section{Characterizing 60GHz UAV Links}
\label{sec:design}
\input{capacity-v5.tex}

\section{{\large{\system}} Design}
\label{sec:design2}
\input{design-v4.tex}
\label{sec:algo}

\section{Implementation and Evaluation}
\label{sec:impl}
\input{evaluation3.tex}

\input{video.tex}

\vspace{-0.15cm}
\section{Large Scale Simulations}
\label{sec:simulate}
\input{simulation.tex}

\vspace{-0.15cm}
\section{Related Work}
\label{sec:related}
\input{related.tex}

\section{Conclusion}
\label{sec:conclusion}
\input{conclusion2.tex}

\newpage
\bibliographystyle{ieee}
 \bibliographystyle{ACM-Reference-Format}
 \bibliography{paper}
 \normalsize
\end{document}

%% file: abstract1.tex
We design and build \system, the first large-scale, autonomous, self-organizing network of unmanned aerial vehicles (UAVs) that are connected using a mmWave wireless mesh backhaul. 
While the use of a {\em mmWave backhaul} paves the way for a new class of bandwidth-intensive, latency-sensitive cooperative  applications (e.g. LTE coverage during disasters, surveillance during rescue in challenging terrains), the {\em network of UAVs} allows these  applications to be executed at operating ranges that are far beyond the line-of-sight distances that limit individual UAVs today. 
\\
To realize the challenging vision of deploying and maintaining an airborne, mmWave mesh backhaul to cater to dynamic applications/events,
\system's design incorporates various elements: (i) Role-specific UAV operations that simultaneously address application tracking and 
backhaul connectivity (ii) Novel algorithms to jointly address the problem of deployment (position, yaw of UAVs) and traffic routing across the UAV network;
and (iii)A provably optimal solution for fast and safe reconfiguration of UAV backhaul during application dynamics.  
We implement \system on 
four DJI Matrice 600 Pros to demonstrate its practicality and performance
through 
autonomous flight operations, complemented by large scale simulations.

%

%% file: intro-v7.tex

\begin{figure}
    \centering
    \includegraphics[width=0.8\columnwidth]{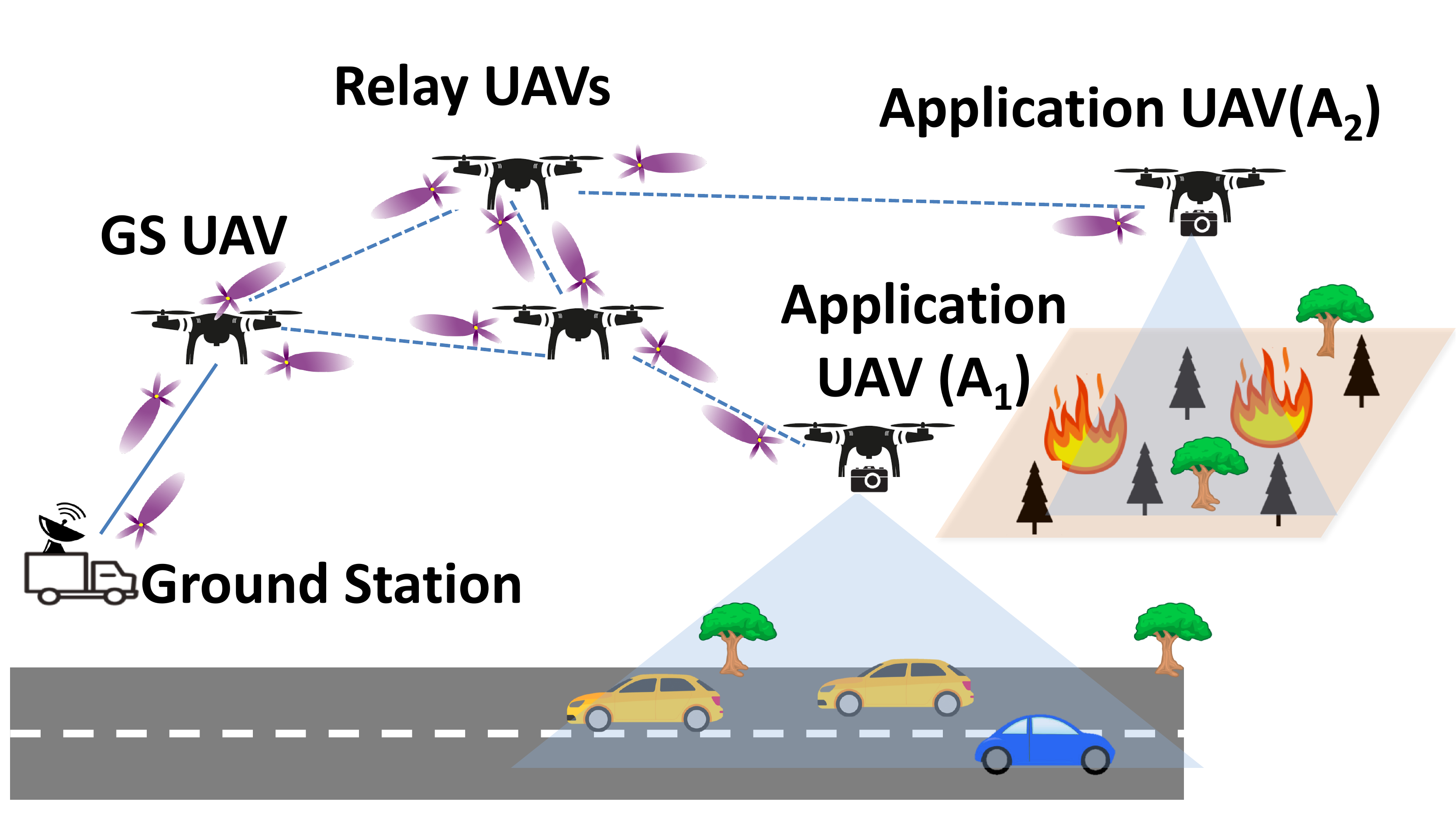}
    \caption{\system deployment using role-based UAV operations.\label{fig:deploy1}}
    \vspace{-24pt}
\end{figure}

We envision a self-organizing network of unmanned aerial vehicles (UAVs),
interconnected through a mmWave mesh backhaul (Fig.~\ref{fig:deploy1}) to enable
a new class of cooperative, \emph{gigabit}~\cite{us-ignite} airborne
applications. Gigabit applications~\cite{utah-ignite} are poised to transform every
aspect of our communities.  With mmWave UAV meshes, we bring the benefits of
ubiquitous Gigabit wireless to a whole new class of applications such as
wide-area high-definition (4K) video streaming; multi-Gbps 5G coverage for users
and first responders in public-safety situations~\cite{skyran,skycore} even
under challenging conditions; on-demand, city-wide IoT connectivity; and
immersive AR/VR streaming from any location~\cite{5g-VR}. We
refer to these applications as \emph{Gigabit Applications in the Sky}, and our
enabling technology as a \emph{Gigabit Network Fabric in the Sky}.


Industry operators have been exploring practical multi-UAV capabilities, from
the ``UAV tower'' proposal by Amazon~\cite{amazon-tower} to the use of high altitude balloons
for internet service by Google~\cite{loon}.  However, these early attempts do
not fully realize the potential of an airborne Gigabit network fabric.
We argue
that the key enabler behind our vision is the capability for multiple UAVs to
exchange/route data and coordinate over a high-bandwidth backhaul to increase
solution coverage to the necessary, practical scale. In this paper, we address
the design and algorithmic challenges behind such a technology enabler for
on-demand deployments, and answer the following question in the
process: \emph{How do we design and deploy an open and flexible
    \underline{network of UAVs} equipped with a \underline{high-bandwidth
    backhaul} for real-time coverage of dynamic events over a wide area?}

\textbf{Key Benefits of a multi-UAV Gigabit Fabric.} 
A multi-UAV, airborne mmWave Gigabit mesh brings with it several key features
for a new generation of applications:

\textit{\underline{Extended Operating Range:}} Existing UAV operations are
lar\-ge\-ly confined to line-of-sight operations that are bounded by the range
of the RF link between the ground station and the UAV.  With a UAV network, the
operating range of all UAVs is extended through the network of UAVs that form a
multi-hop mesh network for communications and control.  Such a network can be
deployed, on-demand to provide LTE coverage in public safety
situations~\cite{skyran} (e.g. areas hit by hurricanes), or as a wide-area
search-and-rescue or surveilliance assistance in inaccessible areas.  


\textit{\underline{High Bandwidth Applications:}} 
Beyond this increased coverage, a Gigabit airborne fabric 
enables high bandwidth applications that are previously impossible.
E.g., Consider UAVs live broadcasting Daytona
500~\cite{foxsports}. Existing approaches are either limited to lower quality
video streams (1080 or 720p), or require a tethered UAV for high-resolution
streams.  With a mmWave (60GHz) mesh backhaul operating at 1Gbps, we can have 
a full-resolution 8K 24FPS live stream using a RED MONSTRO
camera~\cite{redcam} on an untethered UAV.  




\textit{\underline{UAV-Attached Cloud Computing:}} Onboard compute
capability available to current UAVs is highly constrained due to size and power
limitations. With a high-bandwidth, low latency mmWave mesh, cloud-based compute
resources on the ground can be seamlessly integrated.  Real-time, closed loop
UAV network control using data streamed from UAVs and cloud-based AI/analytics
inference applications can now be built over such a Gigabit UAV network fabric.
As an illustration, we enabled real-time object recognition using a 4K video
stream over our 60GHz UAV network. With only a 2$s$ delay between video
capture at the UAV we achieved highly accurate object identification at our cloud
servers with 97\% of supported objects in the video being detected.  





\textbf{Challenges.} Although there are many benefits of a mmWave mesh on UAVs,
realizing it in practice is challenging requiring us to address
multiple systems and algorithmic hurdles.



\begin{figure}
    \centering
    \begin{subfigure}{0.95\columnwidth}
        \centering
        \includegraphics[width=0.8\columnwidth,height=3.5cm]{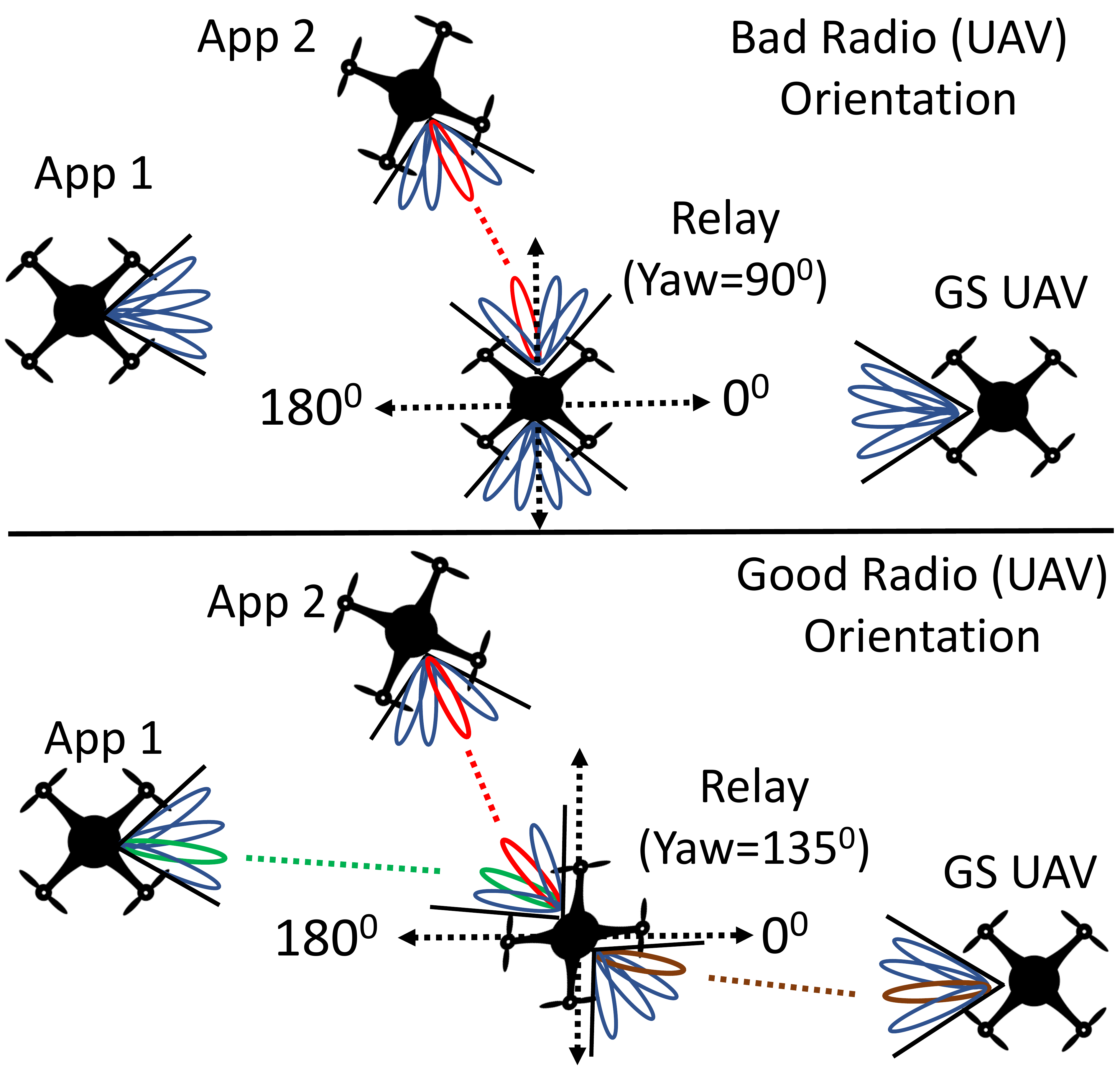}
        \vspace{-0.5cm}
    \end{subfigure}
        \caption{\small{UAV orientation complements 60 Ghz beam selection. Blue -- All possible radio beams. Red, Green, Brown -- Selected beams after successful connection. \label{fig:new_orient}}}
    \vspace{-0.4cm}
\end{figure}

\textit{\underline{High Capacity vs Fragile Connectivity:}} 
60GHz links are
fragile and constant UAV mobility, coupled with multiple hops, exacerbates the connectivity challenge.
Unlike traditional WiFi radios which use omni-directional antennas, the 60GHz
radios use phased-array antennas, each with a limited field-of-view (FOV) a.k.a. sector of operation.
As we highlight later in \S\ref{sec:design}, 
 the achievable throughput is unequal across this entire sector,
with the highest throughput available at $\ang{0}$ beam angles and decreasing as
the beams are steered towards the edge of the FOV. 
Consequently, even with
multiple radios covering $\ang{360}$, the orientation of the
radio with respect to its sender/receiver (as determined by the UAV's yaw -- UAV
orientation along the vertical axis) has a significant impact on
end-to-end performance, particularly during radio capacity sharing in mesh
deployments (e.g. Fig.~\ref{fig:new_orient}). Such an impact (UAV's yaw) on 60 GHz link performance is unique to UAVs,
and is complementary to the well-studied 802.11ad link-layer mechanisms (beam and rate selection) within a single radio at a fixed orientation.

\textit{\underline{Managing the Backhaul Network:}} In order to optimally manage
the backhaul network, one needs to jointly solve for the optimal (a)
\emph{topology} of the UAVs, and (b) \emph{traffic route} between the UAVs and
the ground station. The topology, defined by the number of UAVs deployed along
with the position and yaw of each UAV is tightly coupled with
the routing policy employed throughout the backhaul network: Position and
yaw of each UAV determines its connectivity and capacity to its neighboring
UAVs, which in turn affects how this capacity is shared by multiple end-to-end
traffic flows from other UAVs.  Hence, solving for either topology or routing in
isolation results in a suboptimal UAV deployment solution.  Adding to the
problem's complexity is the need to find a solution using the smallest number of
UAVs, so as to best use the available UAV resources to serve the application.
This challenge, and even subsets of them (e.g. solving for topology discounting
the routing policy) is NP-Hard.  In contrast to prior mesh networking
solutions~\cite{olsr,ett,etx}, this three-dimensional problem  (position,
orientation and routing) is unique to reconfigurable mmWave UAV networks,
thereby requiring a novel approach.  To the best of our knowledge, we are the
first to address it comprehensively.

\textit{\underline{Tracking Event Dynamics:}} With UAV networks, there is a
fundamental tradeoff between the coverage needs of the application or mission,
and the requirements of the UAVs to maintain mesh connectivity. Mobility
requirements to meet app\-li\-ca\-tion goals (e.g. tracking a fast moving vehicle or
event) may result in network disconnection, preventing the mesh from providing
always-on, high-bandwidth connectivity between UAVs and the ground station.
Conversely, maintaining a static configuration to ensure high bandwidth routes
can limit the UAV network from providing the necessary coverage demanded by
applications.  
Further, severe limitations
on power and weight requirements of UAV payloads,  
requires
us to carefully balance the use of multiple mmWave radios (for increased
connectivity/capacity) with that of multiple sensors  such as lidars, cameras, etc.(for better application
coverage).

\textit{\underline{Safe Re-configuration of the Backhaul:}} When catering to
event dynamics, the UAVs need to migrate (move) to a new configuration to
continue their coverage and satisfy application demands. However,
this must be conducted quickly, yet in a \emph{collision-free} manner
-- a challenge that is central to the self-configuring capability of the UAV
backhaul network. 

\textbf{\system.} In this paper, we design and deploy \system, the first
large-scale, airborne, self-organizing, multi-UAV network that operates over a
60GHz mesh backhaul.  With each UAV carrying multiple 60GHz interfaces, the
60 GHz mesh network forms a \emph{data plane} that carries high bandwidth data
between all UAVs, and a ground station bridge that is connected to external
resources.   A separate \emph{control plane} (over 60GHz or public LTE) provides
command and control of all \system UAVs in the network.

To simultaneously cater to the
objectives of application coverage and backhaul connectivity, \system employs\\ 
{\em functionally-specialized} UAV operation as shown in Fig.~\ref{fig:deploy1}:
{\em Application} UAVs focus their mobility decisions to meet
app\-li\-ca\-tion-specific objectives such as target tracking, cellular
coverage~\cite{skyran}, etc.\footnote{Optimizing application UAVs' objective is out-of-scope of this work.}, while {\em Relay} UAVs focus theirs on forming and
maintaining a robust backhaul network to connect the Application UAVs to the
ground station even during event dynamics -- the latter forming the focus of \system. This also has the added benefit of
role-optimized payload design: Application UAVs carry a payload with more
sensors (e.g. cameras, Lidar, etc.) and fewer radios, while {\em Relay} UAVs
carry more radios and fewer sensors.  \textit{Note however that UAVs can be
repurposed for either roles as warranted by the \system's design.} Finally, a
\emph{Ground Station} UAV hovers above or close to the station on the ground,
and serves as an airborne last-hop link connecting \system's network to the
outside world.
\system's key contributions are: 

\textit{\underline{Characterizing 60GHz Links on UAVs.}} 
To understand the impact of UAV's yaw (along with distance) on its link performance and consequently
\system's deployment decisions, we conduct extensive
measurements of UAV-mounted 60GHz links over varying distances and yaw angles.
 A comprehensive, yet practical model of 60GHz link throughput on UAV
platforms as a function of its yaw and distance is determined, to complement the traditional 802.11ad link layer adaptations.  


\textit{\ul{Novel algorithm for efficient positioning, orientation and traffic routing
across UAVs.}} Through a novel two-step optimization across radial and angular
deployment directions, \system \emph{jointly} solves the problem of UAVs network
configuration across three dimensions: (i) the appropriate quantity and position
of UAVs to be deployed; (ii) the yaw of the UAVs such that the position of the
onboard 60GHz interfaces establish a desired mesh connectivity topology between
the deployed UAVs as well as the ground station (GS); and (iii) the routing of
traffic between UAVs and the GS to satisfy the application's traffic demands.

\textit{\ul{Provably optimal algorithm for fast backhaul re-configuration.}}
\system employs a provably safe algorithm that leverages multiple layers of
altitude to ensure that reconfiguration of the UAVs' position and yaw is carried
out quickly (in shortest time) in response to dynamic events, while remaining
collision-free.

\textit{\underline{Real-world implementation of \system.}} We implement \system
on a UAV network of four DJI M600Pro platforms, and evaluate its performance
under real-world conditions.  
Application UAVs carry one Mikrotik 60GHz
interface~\cite{microtik} and a 4K camera, while Relay and Ground Station UAVs
each carry upto three 60GHz interfaces. Routing and managing the UAVs
and their 60GHz radios are done via an on-board Intel Core i7 platform.  The \system
controller runs on AWS~\cite{aws}, and is connected to all UAVs via a LTE
interface on each UAV.  

Through several real-world event tracking experiments, we first demonstrate \system's
ability to  efficiently deploy, maintain and adapt its mmWave backhaul network
in real-time to support both dynamic communication (data transfer) and sensing (live video
streaming and analytics) applications.  Next, using \system's emperical
throughput model, we conduct large-scale simulations to show that \system's
algorithms can support {\em twice} the traffic demands for a given number of UAVs,
or satisfy the desired traffic demands using {\em half} the number of UAVs,
over baseline solutions addressing either of the respective objectives.
A video of \system demonstration can be found at~\cite{videodemo}.

%% file: capacity-v5.tex

Outdoor 60Ghz radios require LOS and
appropriate relative orientation between radios to establish a connection.
Therefore, it becomes critical 
to understand and model the impact of UAV's yaw (orientation) on link performance at coarse time scales (secs),
which is essential for \system's deployment decisions.
Note that this is different from and complementary to traditional mmWave link models~\cite{Torkildson,rappaport,shahsavari}
that capture link performance (e.g., beam selection) on fine time-scales (millisecs), but for a given radio orientation.
%

\begin{figure}
    \centering
    \includegraphics[width=0.8\columnwidth]{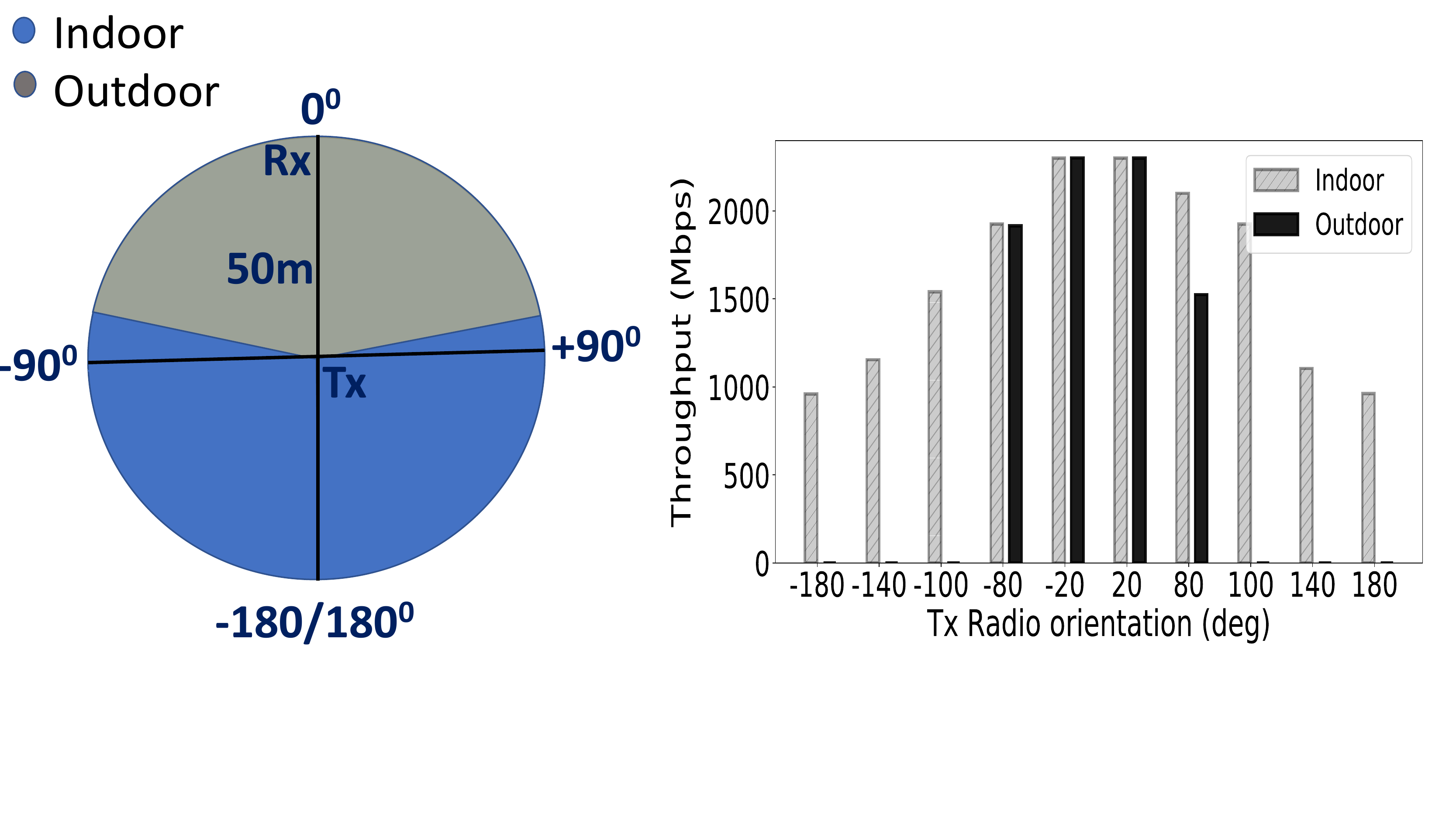}
    \vspace{-1.2cm}
    \caption {Angular coverage \& Link performance  -- Indoor vs. outdoor \label{fig:in_out}}
    \vspace{-0.6cm}
\end{figure}

\textbf{Indoor vs. Outdoor Angular Coverage.} To highlight the impact of multipath-free outdoor operating environment, we compare the performance of our
60GHz platform in outdoor airborne scenarios, against that in a large indoor
hall.  Fig.~\ref{fig:in_out} shows the azimuth angular coverage and throughput,
 when Tx and Rx radios are placed 50m apart LOS, in both indoors and
outdoors. We change the relative angles between the Tx and Rx radios
by rotating the Tx radio about its fixed position. We see that when the radios
are indoors, rich multipath allows coverage to be maintained as long as the Tx radio is pointed
$\pm$\ang{180} of the Rx radio.  However, when the UAVs are moved outdoors, angular coverage
is limited to $\pm$\ang{80} --- a significant drop from that in the indoor
measurement.  

\begin{figure}
    \begin{minipage}[b]{0.48\columnwidth}
        \includegraphics[width=\columnwidth, height=3cm]{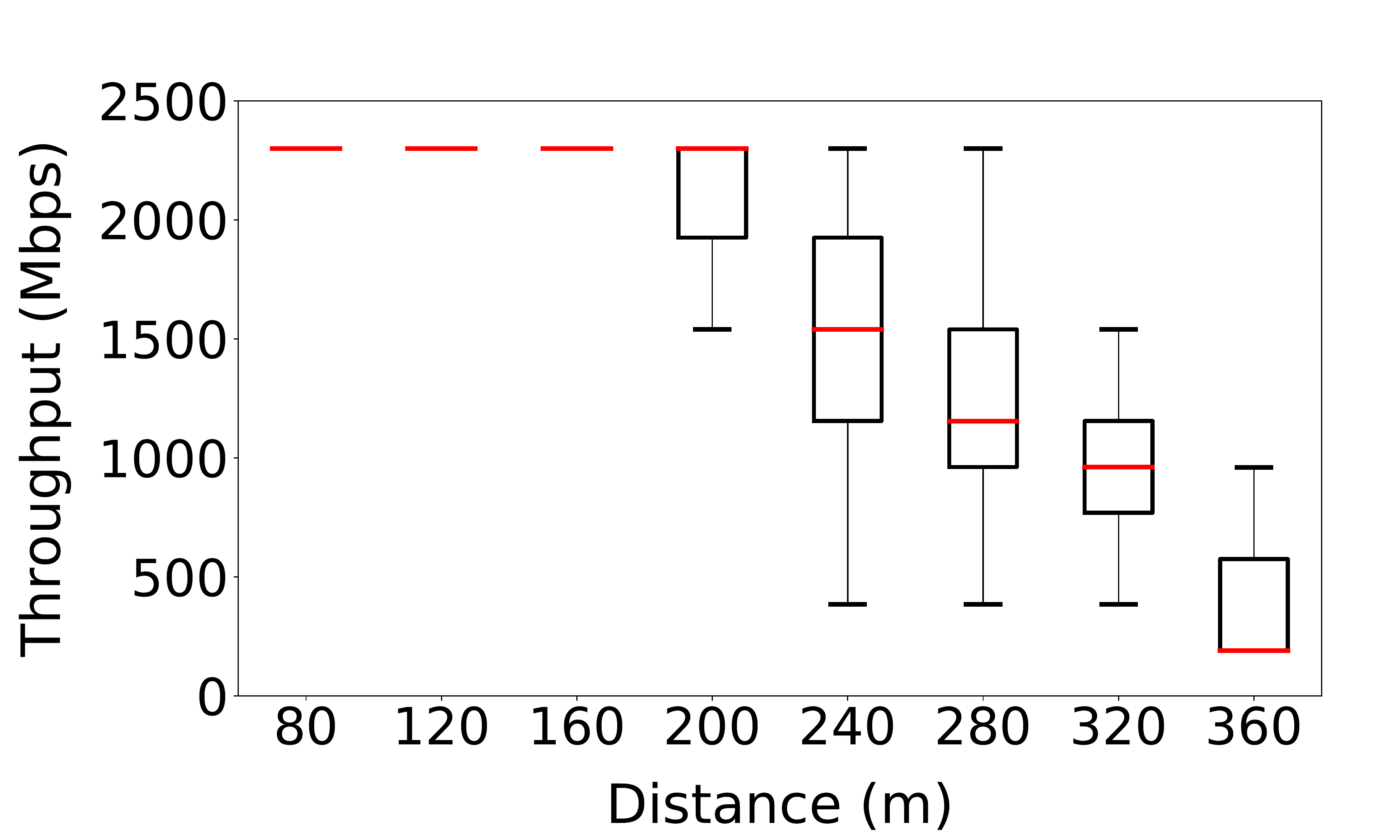}
        \caption {Throughput vs. Distance \label{fig:bitrate}}
    \end{minipage}
    \hspace{2pt}
    \begin{minipage}[b]{0.48\columnwidth}
        \centering
        \small
        \begin{tabular}{||c c c||}
            \hline
            Mcs & RSS & Bitrate  \\
            Index   & (dBm) & (Mbps) \\
            \hline
            8 & -61 & 2310\\
            7 & -62 & 1925\\
            6 & -63 & 1540\\
            4 & -64 & 1155\\
            3 & -65 & 962.5\\
            2 & -66 & 770\\
            1 & -68 & 385\\
            0 & -78 & 27.5\\
            \hline
        \end{tabular}
        \caption{802.11ad bitrate to RSS mapping \label{fig:rss_bitrate}}
    \end{minipage}
    \vspace{-18pt}
\end{figure}


\subsection {60GHz Link Measurement Study} \label{sec:measurement} We now briefly discuss the
details of our  measurement study of the 60GHz radios mounted
on  UAVs.  Further details of our 60GHz operating setup can be found in
\S\ref{sec:impl}.  

\textbf{60GHz Radios.} We mount MikroTik WAP 60G~\cite{microtik} long-range
60GHz radios on DJI M600Pro UAVs for our experiments. These 802.11ad devices
utilize a Qualcomm QCA6335 chipset and a 32 phased array antenna, 
and support three non-overlapping channels. 
Proprietary beam and rate selection algorithms based on the RSS are used to select
optimal operating parameters for each 60GHz link.  A single link supports PHY
bitrate of up to 2.3Gbps (Modulation index 0 to 8). 
In our experiments, we fix the transmission power to maximum, to achieve maximum
link coverage.


\textbf{Data Collection.} We characterize the 60 GHz link performance using UDP
traffic. We fly two UAVs (a Tx and a Rx UAV) at 60m altitude (has a
clear LoS), and increase the separation distance between them from 40 to 400m, in steps of
40m. At each distance, we rotate the Rx and Tx UAVs in turn from \ang{-100}
to $+$\ang{100} (\ang{0} being the perfect alignment of Tx/Rx antennas), while keeping the other UAV
at \ang{0}. We send UDP traffic using \texttt{iperf3} and log the throughput
every second. We also log the PHY bitrate, beam angle, RSS, and UAV telemetry
data (position and yaw).

Link throughput is determined by two main factors -- 
distance and angular difference between the radios. To analyze the effect of each, we decouple them
and study their impact individually.


%
\textbf{Throughput vs. Distance.} 
We fix the angular difference between two UAVs' radios to \ang{0} (\ie both the
Tx and Rx radios are directly facing each other) and measure the impact of
distance on the throughput. Fig.~\ref{fig:bitrate} shows the change in the
received throughput for different distances between the two UAVs.  The high
variance in the  instantaneous throughput is due to the UAV's volatility (e.g. vibrations, jerks) 
in the air, to which, 802.11ad's highly sensitive rate selection mechanism adapts.  
Even a small change of 1dB in RSS results in radios selecting a different PHY rate
as shown in Fig.~\ref{fig:rss_bitrate}.
Nevertheless, we still observe a clear monotonic, declining trend  in
the median link  throughput as the distance increases.

\begin{figure}
    \centering
    \begin{subfigure}[b]{0.48\columnwidth}
        \includegraphics[width=\textwidth]{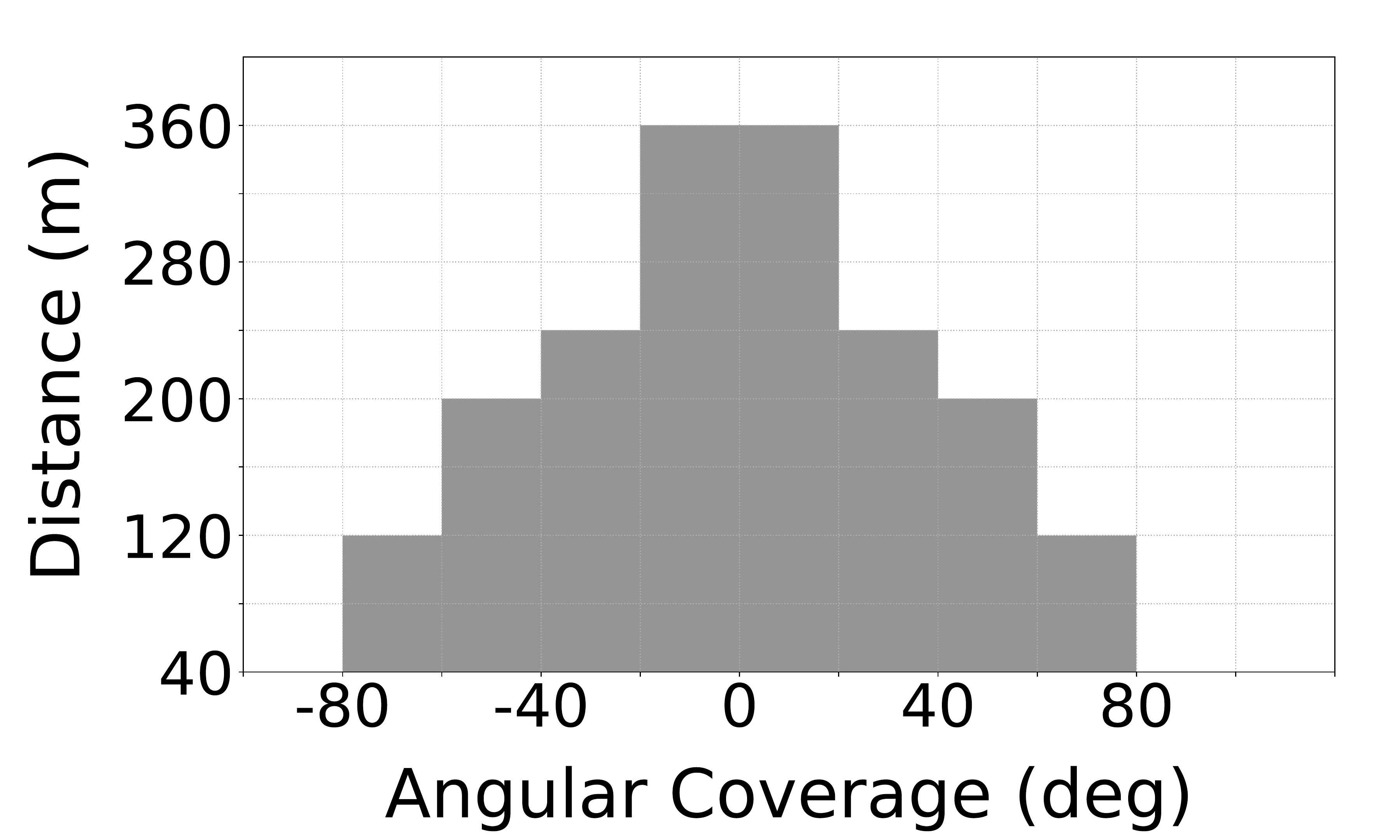}
        \caption{Distance vs. Angular Coverage \label{fig:angle}}
    \end{subfigure}
    \hspace{2pt}
    \begin{subfigure}[b]{0.48\columnwidth}
        \includegraphics[width=\textwidth]{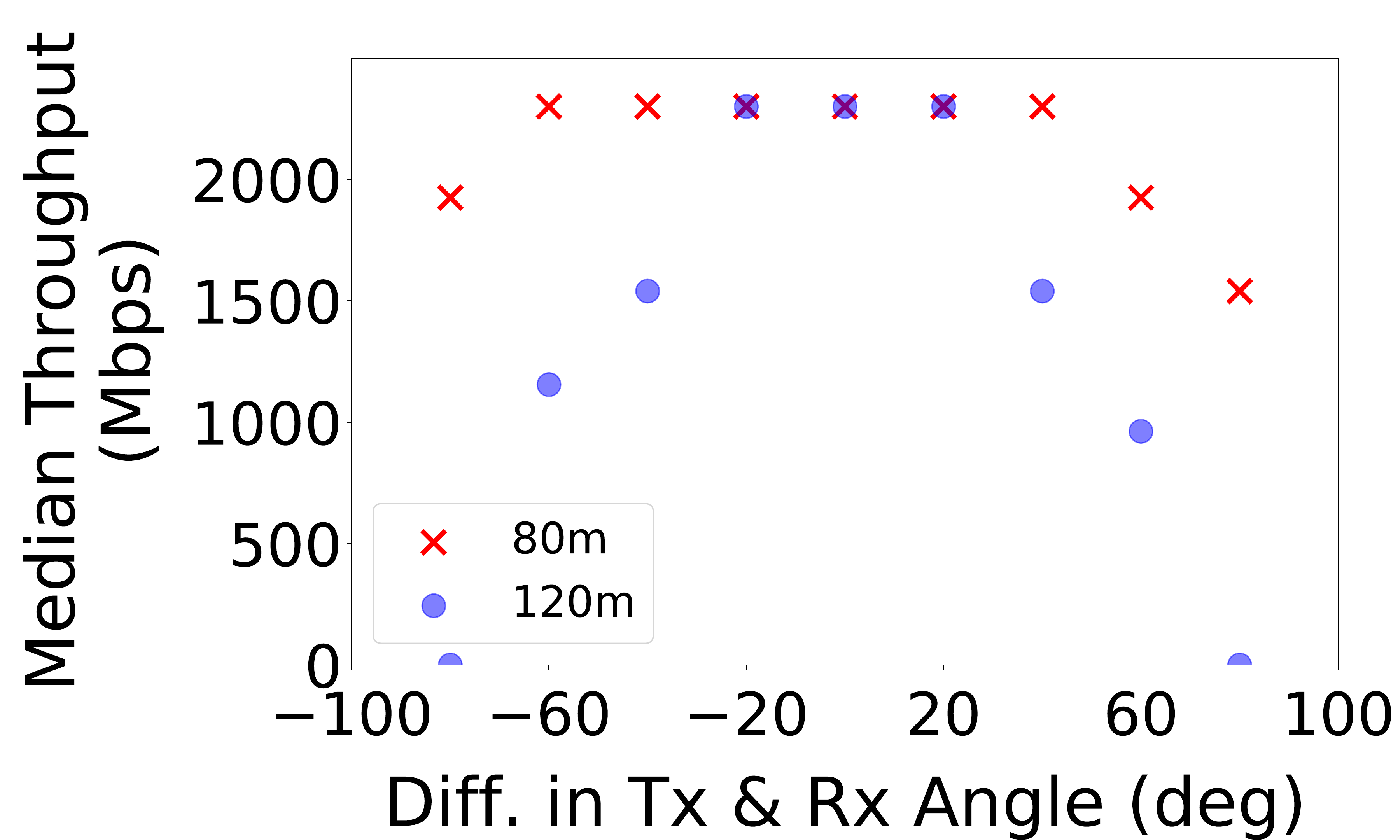}
        \caption {Median throughput for diff. in Tx \& Rx angles, 
            at 80m and 120m  \label{fig:angle_bitrate}}
    \end{subfigure}
    \vspace{-0.3cm}
    \caption{Impact of UAV yaw}
    \vspace{-0.3cm}
\end{figure}
 
\textbf{Throughput vs. Relative Yaw.}
For a given yaw angle (\ang{0}) of the Rx UAV, we vary the relative yaw angle of the Tx UAV
from $-$\ang{100} to $+$\ang{100}, while sampling throughput measurements at
regular yaw intervals. 
%
%
Fig.~\ref{fig:angle} shows a binary map denoting
successful 60GHz link connectivity (i.e. when throughput > 100Mbps)
at different distances and relative yaws.
Although the radio's (MikroTik) specifications prescribes an antenna coverage of only
\ang{40} ($-$\ang{20}to $+$\ang{20}), we observe that connectivity can be
established across relative yaws from $-$\ang{80} to $+$\ang{80} when the UAVs
are \SI{80}{\metre} apart, while only between $-$\ang{20} and $+$\ang{20} for distances
greater than \SI{240}{\metre}.  This 
is because the side-lobes of the Tx beam have sufficient energy to establish connectivity
at short ranges, increasing the angular coverage. 
However, the side-lobes have 
no impact at longer ranges, where the main lobes determine coverage.
This interplay between main and side-lobes is also reflected in the 
achieved throughput over the 60GHz channel:
Fig.~\ref{fig:angle_bitrate} shows that at short
distances (\SI{80}{\metre}), the maximum median throughput is achieved
over a wider range of relative yaw angles 
than at longer UAV distances (\SI{120}{\metre}).


\begin{figure}
    \centering
      \includegraphics[width=0.8\columnwidth]{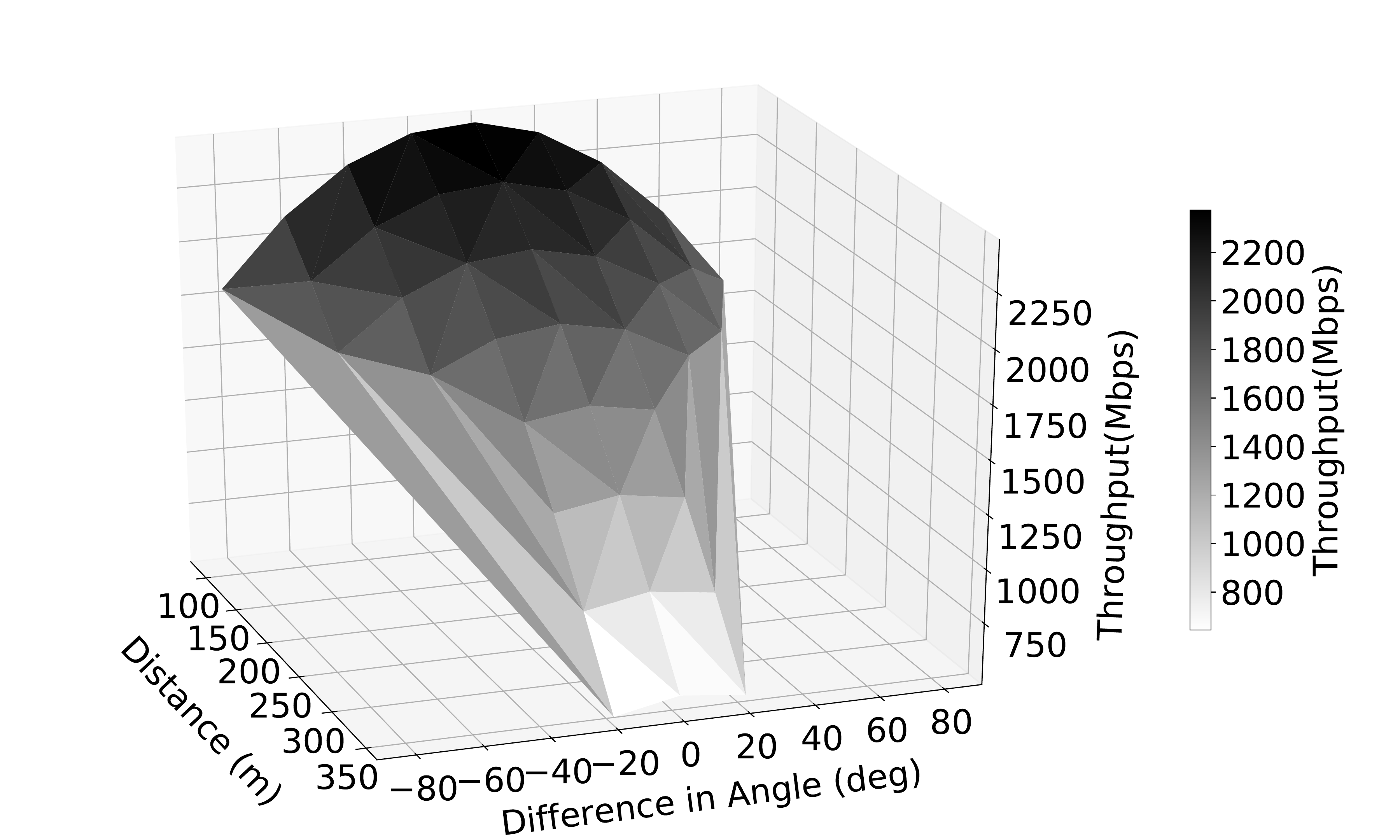}
    \vspace{-0.30cm}
    \caption {Surface-curve fitting for the throughput model.\label{fig:curve_fit}}
    \vspace{-0.45cm}
\end{figure}

\begin{figure}
    \centering
        \includegraphics[width=\columnwidth,height=3.5cm]{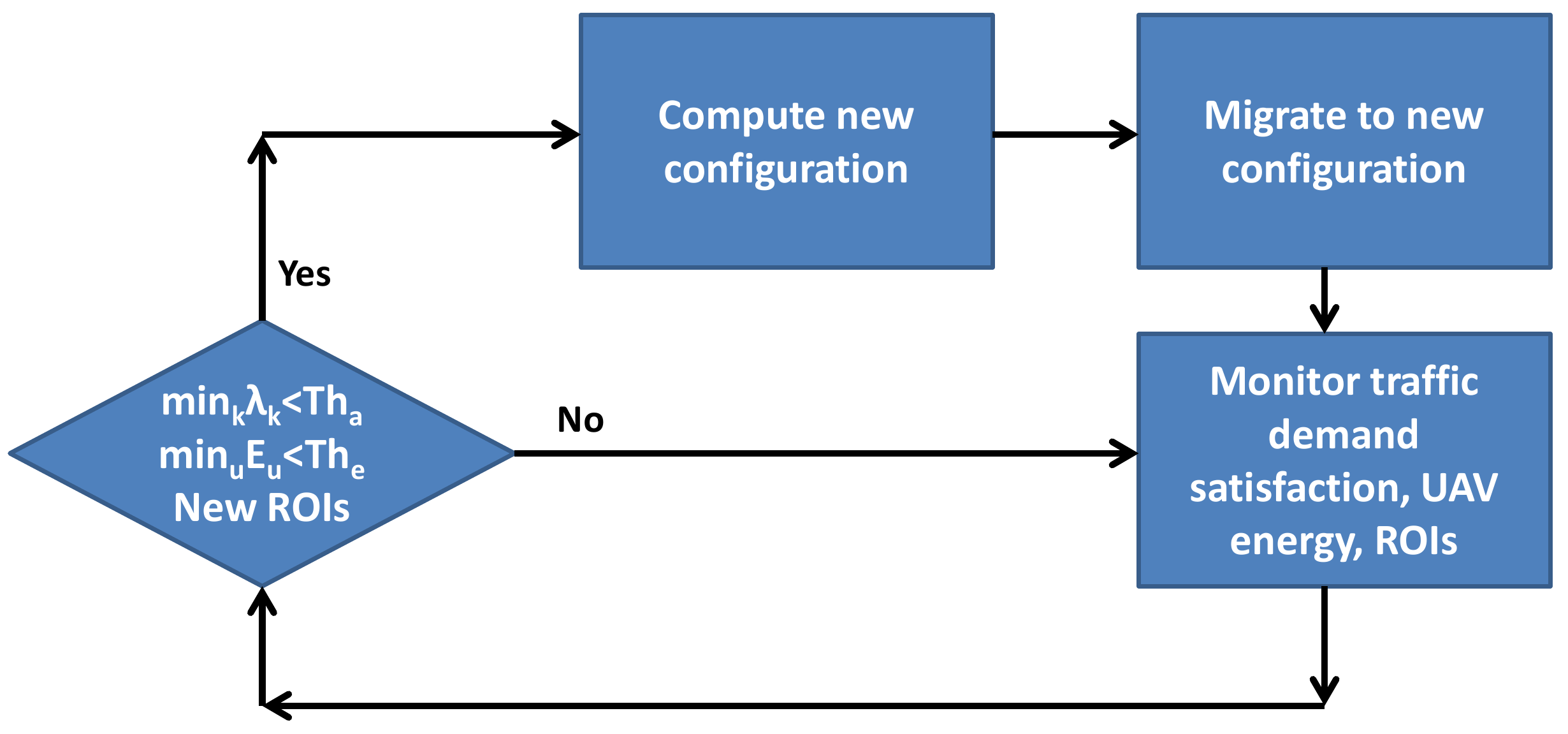}
    \vspace{-0.5cm}
        \caption{\system operational sequence.\label{fig:operation}}
    \vspace{-0.45cm}
\end{figure}




\subsection{60GHz UAV Link Throughput Model} 
Based on our measurement study, we can now characterize the throughput model of our 60 GHz UAV links as a function
of both the relative UAV distance ($\mathbb{D}$) and yaw angles ($\Delta \phi$). 
A weighted least-square regression approach that accounts for
heteroscedasticity of the throughput numbers in our dataset, results 
in the following quadratic  model as the best fit: $\mathbb{C} = a.\mathbb{D}^2 + b.(\Delta \phi)^2 +
c.(\Delta \phi).\mathbb{D} +d.\mathbb{D} +e.(\Delta \phi) +f$, where $a,b,c,d,e,\; \mbox{and}\; f$ are  constants derived from the regression curve-fitting technique. 
Fig.~\ref{fig:curve_fit} shows the surface curve-fitting of the model that accurately captures the parabolic dependence of throughput on both 
$\mathbb{D}$ and $\Delta \phi$.

\textbf{Leveraging the model.}
While a throughput's quadratic dependence on distance is expected in a LoS environment,
the model's contribution lies in its incorporation of the joint dependence on both relative distance and yaw.
Indeed, while 802.11ad's link adaptation mechanisms are meant to handle instantaneous (millisecs granularity)
throughput fluctuations (due to UAV volatility), \system's throughput model is meant to {\em complementarily} capture
the impact of UAV's yaw on first-order (median, mean) throughput statistics. The latter allows \system to optimize its
deployment decisions that are executed at coarse time scales (mins granularity). 
Further, constructing this model is a one-time effort that depends only on the mmWave radios and antennas employed.

%% file: design-v4.tex
\subsection{\system Overview}
\system envisions to provide real-time coverage and tracking  of a target
application (e.g. mobile connectivity, live video surveillance, etc.) from a
Ground Station (GS) that could be far away, using one or more UAVs. 
To allow applications to 
flexibly and dynamically define their  ``Regions Of Interest" (ROIs) on the ground based on their individual requirements, \system decouples the objectives of application coverage and backhaul connectivity, while focusing on the latter.
It realizes this through {\em functionally-specialized} UAV operation:
{\em Application} UAVs  are tasked with specific ROIs by the application and move freely within their AOI (``airspace of interest"; i.e., airspace above the ROI assigned) to optimally track the target application, while {\em Relay} UAVs
focus only on  forming, maintaining and adapting a
robust backhaul network to connect the dynamic Application UAVs to the GS.
This has the added benefit of role-optimized payload design: Application UAVs carry a payload with more sensors (e.g. cameras, Lidars, etc.) and few radios, while {\em Relay} UAVs carry more radios and less sensors. 
Note however, UAVs can be repurposed for either roles as 
required by the application. 

With the application UAVs being directed by the application,  \system focuses on orchestrating the backhaul network of Relay UAVs to satisfy the application requirements.
 \system
accomplishes this through a simple operating control loop, as shown in
Fig.~\ref{fig:operation}.  When the \system network is first initialized or a
configuration update is triggered, it \emph{(i) determines an efficient network
configuration} of Relay UAVs to meet the desired application objectives.  Then, \system
\emph{(ii) computes a seamless migration plan} to enable UAVs (both Relay and Application) to optimally and
safely deploy and reconfigure themselves in this configuration.  Once the UAVs are in their
new configuration, \system continuously \emph{(iii) monitors performance} across
the 60GHz mesh and triggers a
reconfiguration step (along with migration, if necessary) if: (a) traffic satisfaction (of application sessions,
$\lambda_s$) falls below a desired threshold ($Th_a$), (b) UAVs need to be brought back 
(and replaced with another UAV) for energy (battery/gas) 
replenishment (when energy below $Th_e$), or (c) when Application UAVs have been re-tasked to cover new ROIs.

{\bf Remark:} Note that \system does not focus on MAC/PHY optimizations like beam and rate selection on links. 
Its coarse time-scale network optimizations are complementary to such fine time-scale link-layer 
optimizations that can be leveraged from prior art~\cite{shahsavari,shahsavari2019,haider:2016}.

\subsection{Determining Efficient Configurations}
\label{subsec:config}
\vspace{-0.5cm}
\subsubsection{Understanding the Problem} 
\textbf{Formulation.} 
\system's objective is to cover and track an
application/event that spans across multiple regions on the ground (ROIs, refer Fig.~\ref{fig:deploy1}), from corresponding, non-overlapping zones above in the sky (AOIs, $\mathcal{A}_k;\; k\in
\mathcal{K}$) at
a given altitude (i.e. a 2D deployment plane\footnote{The benefits of deploying UAVs at different altitudes (other than during transient mobility) are unclear, especially in light of connectivity requirements that are better addressed at the same altitude.}).  Thus, each of these zones generates a traffic
demand $T_k$ (e.g. video streams) that needs to be delivered through one or more Relay UAVs
to the ground station UAV (node $0$, positioned above GS, last-hop connecting the Relay UAVs to the GS), which is assumed to be at the origin in the
deployment plane. The static GS UAV that anchors the backhaul UAV-relay network, has one of its radios pointed down towards the GS (Z axis), unlike all other UAV radios (oriented in the XY-plane).

This problem of Relay-UAV network configuration (DCR) requires a {\em joint} optimization across three components: (i) {\em Deployment} (placement) of UAVs; (ii) {\em Connectivity} between UAVs and ground station (GS) as determined by orientation (yaw) of UAVs' radios; and (iii) {\em Routing} of traffic between UAVs and GS to satisfy the application's traffic demands.
 For easier exposition of the formulation and solution, a
single ground station (GS) is considered. 
Furthermore, from a practical realization standpoint, we consider flow routing
to be non-splittable, i.e. traffic in a session between Application UAVs and the GS 
must be routed without being split at any intermediate UAV, albeit, the latter can route flows of multiple sessions. 
The formulation and solution can be adapted to meet the setting of
multiple ground station UAVs and splittable flow easily.

\vspace*{-0.4cm}
\small
\begin{eqnarray}
\mbox{DCR:}&  \mathcal{D}^*=\min_{\mathcal{D}} |\mathcal{D}| & \\
\mbox{s.t.,} &\sum_i x_{ki}^k = 1,& \forall k\in \mathcal{K} \nonumber \\
&\sum_i x_{i0}^k = 1,& \forall k\in \mathcal{K} \nonumber \\
&\sum_v x_{vi}^k = \sum_u x_{iu}^k, & \forall
    (i,k):i\neq \mathcal{K} \cup \{0\} \nonumber \\ 
&\sum_{u}\gamma_{iu}s_{iu}^m + \sum_{v}\gamma_{vi}s_{vi}^m \le 1,& \forall i, m\in[1,M] \nonumber \\
&\gamma_{iu}C_{iu} \ge \sum_k \lambda T_k x_{iu}^k,& \forall u, i \in \mathcal{D}\nonumber \\
&\gamma_{vi}C_{vi} \ge \sum_k \lambda T_k x_{vi}^k,& \forall v, i \in \mathcal{D} \nonumber 
\end{eqnarray}
\normalsize

$\mathcal{D}$ denotes the network configuration to be determined, $\mathcal{D} =
\{\cup_i(p_i,\phi_i,X_{i}) \}$, where each element in the configuration
corresponds to that of a UAV $i$, namely its position $p_i$, orientation
$\phi_i$ and its flow routing $X_i = \{x_{iu}^k\},\; \forall u\neq i,k$. Each
element of $X_i$ is a binary variable corresponding to whether flow $k$ is
routed from $i$ to $u$ (i.e. $x_{iu}^k$).  $C_{uv} = Cap(p_u,p_v, \phi_u,\phi_v)$
is the expected throughput of the link $u-v$, which (from \S\ref{sec:design}), depends on the distance between them
(i.e their locations, $p_u,p_v$) 
and their respective radios' orientation 
($\phi_u,\phi_v$), i.e. $Cap(p_u,p_v,\phi_u,\phi_v) = Cap(\Delta
r(p_u, $\\$p_v), \Delta \phi(\phi_u,\phi_v))$. 
The adjacent radios (total of $M$) at a UAV are separated by  $\frac{2\pi}{M}$ radians, with each radio responsible
for covering its logical sector of $\frac{2\pi}{M}$ radians.
 $s_{iu}^m$ is a binary variable that captures the
presence of UAV $u$ in the $m$-th sector of UAV $i$ and depends on their
respective positions $p_i,p_u$ and also UAV $i$'s orientation $\phi_i$. Note
that a UAV's position $p_u$  is captured in terms of its polar coordinates
($p_u= (r_u,\theta_u)$) taken with respect to the GS UAV at origin. 

{\bf Objective.} In practical environments, there is a need to maintain additional drones to meet increasing application demands or to replace existing drones with depleting energy. Hence, DCR's objective is to find the 
deployment configuration $\mathcal{D}^*$ with the least number of Relay UAVs (cost) that can 
completely satisfy ($\lambda=1$) the current traffic demands, while yielding a feasible flow routing that accounts for directional
connectivity/capacity constraints.  However, its flexible framework 
{\em allows for selecting appropriate configurations that maximize demand satisfaction for a given deployment cost as well} (by evaluating solutions for varying $\lambda \in (0,1]$).  
 The first three constraints above correspond to flow origination (at UAVs
covering application $k$), termination (at GS UAV $0$) and conservation
(UAVs in between). Note that \system can handle bi-directional traffic as
well, where upstream and downstream flows will be considered as separate
sessions.  The final three constraints are responsible for coupling flow routing
to the notion of directional connectivity and capacity at each of the UAVs --
specifically, for all the links (incoming and outgoing) that share the capacity
of a radio (i.e., all the neighbors that fall in the sector of that radio $m$) at
a UAV $i$, there must be a feasible time-sharing ($\gamma_{iu},\gamma_{ui}$)
of those links (third last constraint), taking into account their respective
capacities, that is capable of carrying the desired fraction ($\lambda$) of traffic demands (last two
constraints). \noindent{\bf Energy.} Note that energy consumption in drones is dominated by their mechanical operation rather than connectivity and hence not incorporated in DCR\footnote{Gas powered drones can operate for a few hours~\cite{uav4}}. Nevertheless, \system will have the mechanisms needed to cope with the impact of energy-related network dynamics (Fig.~\ref{fig:operation}).

\textbf{Challenges.} Addressing DCR encounters challenges at multiple levels.
(i) {\em Hardness of computing configurations:} The 3-way optimization in DCR  is unique to the UAV environment and has not been addressed before. Its extremely challenging nature is evident from the NP-hardness of solving even just one component (deployment using Euclidian Steiner trees~\cite{chen:2000},
without considering orientation and routing) or two components (routing and orientation 
assuming deployment is given) jointly. 
(ii) {\em Practical realization and adaptation of configurations:} The Relay UAVs have to periodically adapt and reposition themselves  in a new network configuration to seamlessly track a spatially evolving event (e.g., forest fire) and hence the Application UAVs. Such a migration of network configurations involving multiple UAVs simultaneously needs to be executed both quickly and safely (i.e., avoiding any UAV collisions). 

Given the hardness of the simplest variants of the DCR
problem, 
\system aims to design a solution that
organically integrates the components of deployment, connectivity and routing to
compute efficient configurations, along with the dynamic mechanisms needed to adapt and realize these configurations safely in practice.  

\begin{figure}
\centering 
\begin{subfigure}{0.48\columnwidth}
 \centering
        \includegraphics[width=\columnwidth]{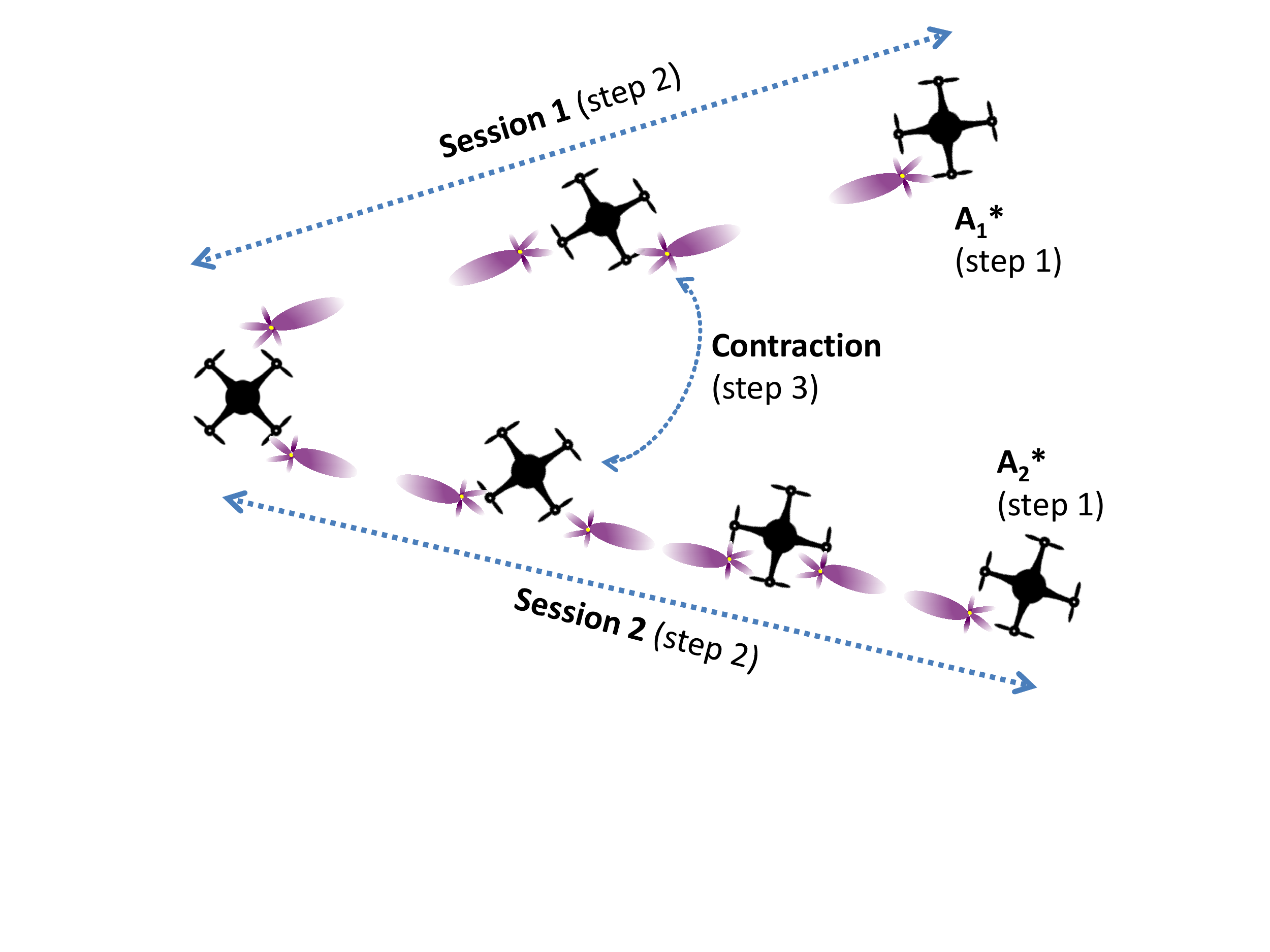}
        \caption{Optimize deployment per-session (radially)}
\end{subfigure}
\hspace{2pt}
\begin{subfigure}{0.48\columnwidth}
 \centering
        \includegraphics[width=\columnwidth]{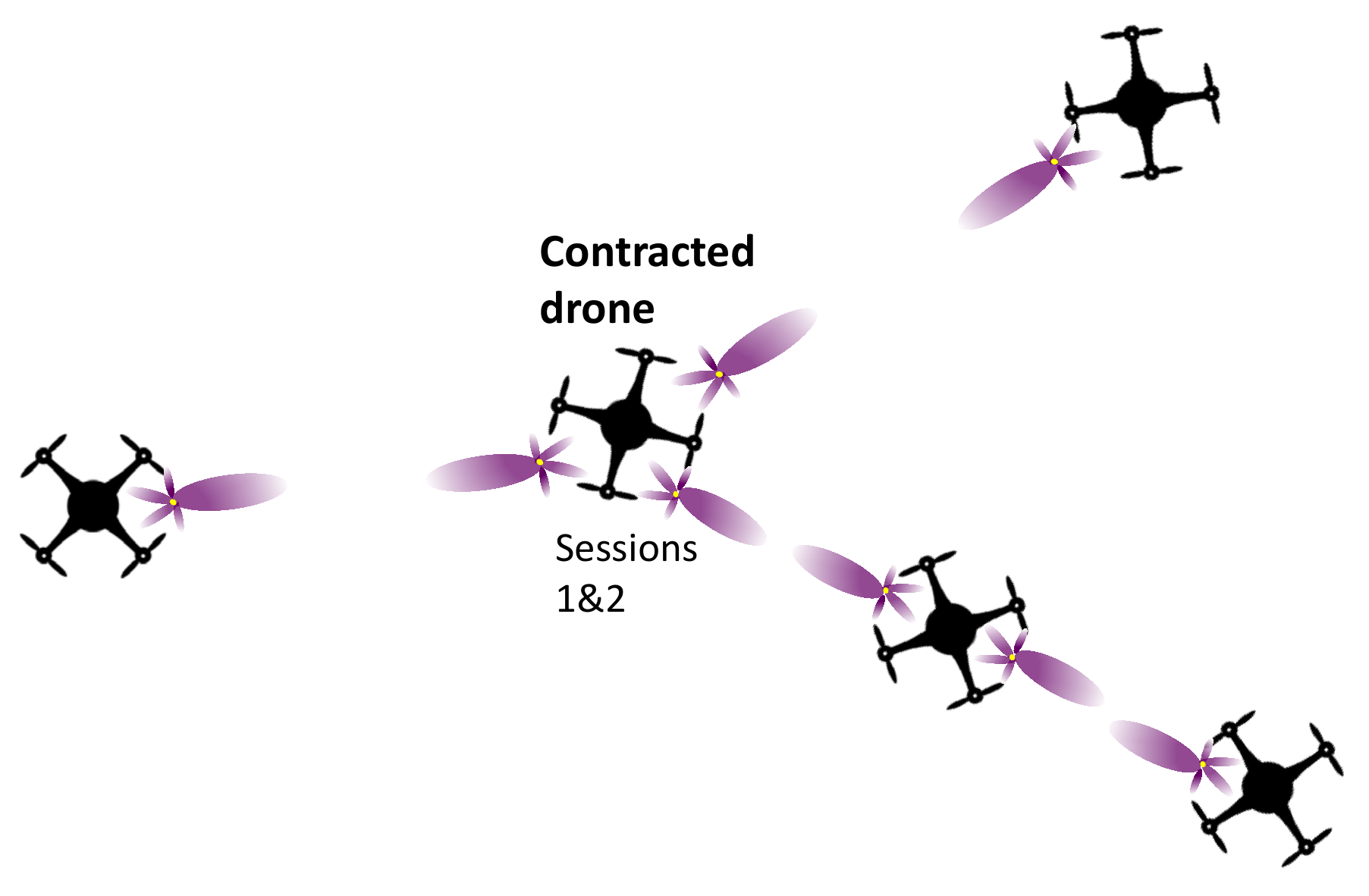}
        \caption{Optimize deployment across sessions (angularly)}
\end{subfigure}
\vspace{-0.2cm}
\caption{Two-step optimization for backhaul deployment.\label{fig:2stepopt}}
\vspace{-16pt}
\end{figure}

  \begin{algorithm}[tbh] 
    \small
    \caption{Joint Deployment, Connectivity and Routing in \system}\label{alg:skyhaul}
    \begin{algorithmic}[1] 
    	\State \% Load-balanced deployment: Radial optimization per session
        \State Input: GS UAV at $p_0 = (0,0)$, Application UAVs at $p_k=\mathcal{A}_k^*=(r_k,\theta_k),\; \forall k\in \mathcal{K}$   
        \State Output: $\mathcal{D}_m^{cur}= \{\cup_i(p_i,\phi_i,X_{i}) \}$ \% backhaul network configuration
        \For{$\phi = [0,\frac{2\pi}{M})$}
        		\For{$m=[1,M]$}
			\State \%app UAVs in GS UAV's sector $m$, orientation $\phi$
			\State $S_{m,\phi} = \{k\in \mathcal{K}:\mbox{sector}(k,\phi)=m\}$ 
			\State $[N_{k,m,\phi},\mathcal{D}_{m,\phi}] = \mbox{Get\_RadialOpt\_Dep}({m,\phi})$
		\EndFor
		\State $\mathcal{D}_{\phi} = \cup_m \mathcal{D}_{m,\phi}$
        \EndFor
        \State $\phi_{GS} = \arg \min_{\phi} \{ N(\mathcal{D}_{\phi}) = \sum_m \sum_{k\in S_{m,\phi}}N_{k,m,\phi}\}$
        \State $\mathcal{D}^{cur} = \mathcal{D}_{\phi_{GS}}$
        \State
        \State \% Contraction: Angular Optimization across sessions
        \For{$m\in[1,M]$}
        		\State \% Initialize all contraction variables
		\State $Con(i,j) = 1,\; \forall i,j\in \mathcal{D}_m^{cur}: \{ S(i)\neq S(j)\; \&$ 
		\State $\qquad\qquad i,j\neq k, k\in S_{m,\phi_{GS}}\}$; and $0$ otherwise
		\Do 
			\State $I_{con} = \mbox{False}$
			\For{$(i,j)\in \mathcal{D}_m^{cur}: Con(i,j)\neq 0$}
				\State $L(i,j) = \mbox{Get\_ContractPoints}(i,j,\mathcal{D}_m^{cur})$
				\State $\hat{L} = \{\ell \in L(i,j): $
				\State $\qquad\qquad \exists\; \mbox{Feasible\_Orient}(i,j,\ell,\mathcal{D}_m^{cur}) \}$
				\State $\ell^* = \arg \min_{\ell \in \hat{L}} \{\sqrt{\sum_{k'\in S(i)\cup S(j)} \Delta r^2(p_\ell,p_{k'})}  \}$
				\If{$\hat{L} \neq \emptyset$}
					\State $\mathcal{D}_m^{cur} \gets \mbox{Update\_Config}(i,j,\ell^*,\mathcal{D}_m^{cur})$
					\State $I_{con} = \mbox{True}$; break
				\Else 
					\State $Con(i,j) = 0$
				\EndIf
				
			\EndFor
		\doWhile{$I_{con}==\mbox{True}$}
        \EndFor
	\State \% Contracting UAVs across adjacent sectors
	\State Repeat Contraction steps, while replacing $\mathcal{D}_m^{cur}$ with $\mathcal{D}_{m,m+1}^{cur} = \mathcal{D}_m^{cur} \cup \mathcal{D}_{m+1}^{cur}$
\end{algorithmic}
\end{algorithm}
\normalsize

\subsubsection{An Efficient Dual-Step Algorithm}
Given the position of the application UAVs, i.e. $p_k = \mathcal{A}_k^*,\; \forall k\in \mathcal{K}$, \system determines an efficient UAV (backhaul) network configuration $\mathcal{D}^*$ to satisfy their traffic demands through a novel two-step optimization (steps 2,3 in Fig.~\ref{fig:2stepopt}): {\em radial} optimization per session, followed by {\em angular}
optimization across sessions. 

\noindent {\bf Radial optimization per session:} 
Given the optimal position of the application UAVs $\mathcal{A}_k^*, \forall k$ with respect to event coverage, \system first determines the GS UAV's orientation that {\em radially} optimizes the deployment of relay UAVs to support the desired traffic demands to/from each session (application UAV) in isolation (steps 1-13, Alg.~\ref{alg:skyhaul}), while load balancing across the capacity of GS UAV's radios.
  It accomplishes this by determining the minimal relay UAV deployment needed for a given GS UAV's orientation (steps 5-10, Alg.~\ref{alg:skyhaul}), and then picking that orientation that yields the minimum deployment among all its orientations (step 12, Alg.~\ref{alg:skyhaul})).
When the GS UAV's orientation ($\phi$) is fixed, this geometrically partitions ($S_{m,\phi}$) the application UAVs, i.e. load balances their traffic demands across the capacity of various radios (sectors, $m$) at the GS UAV (step 7, Alg.~\ref{alg:skyhaul}). In each sector of the GS UAV $m$, \system determines the minimal UAV deployment (step 8, Alg.~\ref{alg:skyhaul}) needed to support the traffic demands of the application UAVs in that sector (i.e. $S_{m,\phi}$) by solving the following problem. 

\noindent \underline{${\mbox{Init\_Deploy}(m,\phi,S_{m,\phi}):}$}
\vspace*{-0.2cm}
\small
\begin{eqnarray}
\mbox{Minimize}_{h_k} &\{\sum_{k\in S_{m,\phi}}N_{k,m,\phi} = \lceil\frac{r_k}{h_k}\rceil \} &\nonumber\\ 
 \mbox{subject to,}&  \gamma_k \cdot Cap(0,(h_k,\theta_k),\phi,0) \ge T_k, \forall k\in S_{m,\phi}& \nonumber \\ 
& \sum_{k\in S_{m,\phi}}\gamma_k \le 1 & \nonumber
\end{eqnarray}
\normalsize

This optimization helps determine the minimal number of UAVs and their deployment positions needed  to support the traffic demand of each session in isolation (without sharing relay UAVs across sessions), while accounting for time  ($\gamma_k$) sharing the same radio's capacity (at the GS UAV) between sessions in the same sector.  With our quadratic  capacity function (Section~\ref{sec:design}),  this can be solved through an equivalent convex optimization (that relaxes ceiling function first, solves the resulting convex problem, then rounds the fractional solution).
Once the UAV deployment is determined, their orientations are identified to deliver the data rates needed to satisfy the corresponding session's traffic demand. 
The union of  the relay UAV deployments  in each sector provides the initial backhaul network configuration that is radially optimized to support the given traffic demands.  

\noindent {\bf Angular optimization across sessions:} 
\system then optimizes the deployment {\em angularly} by contracting (eliminating) relay UAVs with abundant radio capacity,
and re-positioning the remaining relay UAVs so as to efficiently utilize their available radio capacity across sessions, all the while continuing to support the desired traffic demands (steps 16-35, Alg.~\ref{alg:skyhaul}).
During this  optimization in each sector, \system 
checks every pair of relay UAVs that belong to different sessions, whether they can be contracted, i.e., removed and replaced by a single UAV  in a new, appropriate position and orientation, while supporting the aggregate traffic to both the original UAVs (steps 22-33, Alg.~\ref{alg:skyhaul}). When such a pair exists, the contraction is done and the two UAVs are replaced by the  new UAV and the network configuration is updated (steps 27-30, Alg.~\ref{alg:skyhaul}); 
i.e., the neighborhood connectivity for the new UAV along with the routing of flow on its edges, as well as its (and neighbors') orientation are updated. 

\underline{\em Contraction:} Determining whether a pair of relay UAVs can be contracted ($Con(i,j)$) is a challenging problem in itself, given the infinite combinations of position and orientation possible for the new contracted UAV. \system accomplishes this efficiently in two steps: First, it identifies the largest contraction region possible by relaxing the constraint of orientation (i.e., assumes perfect orientation, $\Delta \phi =0$); then, selects a small number of boundary points of this region as potential contraction points (step 23 in Alg.~\ref{alg:skyhaul}). 
Finally, for each of these contraction points, it determines if there exists a feasible orientation for both the new contracted UAV as well as all its neighbors such that existing traffic demands can be satisfied (steps 24, 25 in Alg.~\ref{alg:skyhaul}). 

\begin{figure}
    \centering
    \includegraphics[width=0.7\columnwidth]{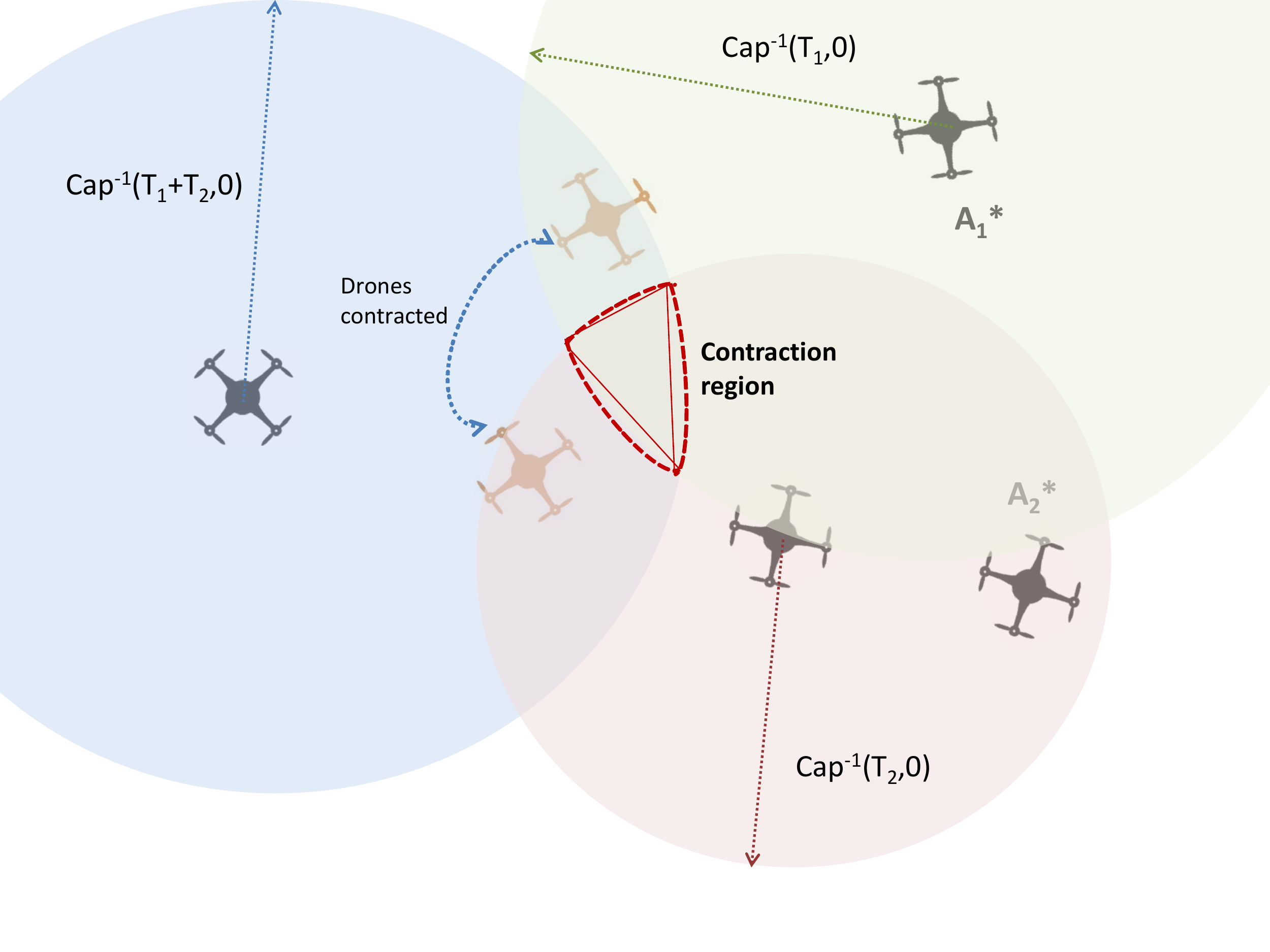}
    \caption{{Bounding feasible region for contraction.\label{fig:contract}}}
\vspace{-0.5cm}
\end{figure}

The contraction region is determined by the polygon formed from the intersection of the circles -- one for each neighbor of the pair of UAVs being contracted and whose radius corresponds to the maximum distance ($Cap^{-1}(T)$) that supports the traffic demand  between that neighbor and the two UAVs as shown in Fig.~\ref{fig:contract}. 
The corner points of this polygon along with its centroid are selected as a representative set of contraction points to be evaluated for a feasible orientation for routing.
If multiple contraction points are feasible, then the one that is farthest from the GS UAV and closest to the application UAVs is selected (step 26 in Alg.~\ref{alg:skyhaul}); with contraction proceeding radially from the GS UAV, this increases the scope for subsequent optimization/contraction.  
In checking whether a feasible orientation exists (for demand satisfaction, $\lambda_{\phi_\ell}^* \ge 1$) for the new contracted UAV and all its neighbors, 
a simple optimization $\mbox{Solve\_Orient}()$ 
is solved for each of the UAVs under consideration.

\noindent \underline{${\mbox{Solve\_Orient}(\ell,B(\ell),T(\ell),\phi_{B(\ell)}):}$}
\vspace*{-0.1cm}
\small
\begin{eqnarray}
&\mbox{Maximize}_{\gamma,\phi_\ell}\qquad \min_{m_\ell}\lambda_{m_\ell,\phi_\ell} &\nonumber\\ 
 &\mbox{\bf{s.t.},}\;   \gamma_{b,m_\ell} \cdot Cap(p_\ell,p_b,\phi_{\ell},\phi_b) \ge \lambda_{m_\ell,\phi_\ell}\cdot T_b,\; \forall b\in S_{m_\ell,\phi_\ell},\; \forall m_\ell & \nonumber \\ 
&\qquad  \sum_{b\in S_{m_\ell,\phi_\ell} }\gamma_{b,m_\ell} \le 1,\; \forall m_\ell & \nonumber  \\
&\mbox{where,}\qquad  S_{m_\ell,\phi_\ell} = \{b\in B(\ell):\mbox{sector}_\ell(b,\phi_\ell)=m_\ell\}&\nonumber
\end{eqnarray}
\normalsize

The above optimization selects an orientation for UAV $\ell$ and a time sharing solution ($\gamma$) for each of its sectors such that  
the maximum amount of its traffic demands can be satisfied. 
For a given orientation of the UAV ($\phi_\ell$),  this optimization can be decoupled across the UAV's sectors ($m_\ell$), where it becomes linear and can be solved individually (optimally) in each sector. Thus, the above optimization has the following closed-form solution.

\vspace*{-0.4cm}
\small
\begin{eqnarray}
\phi_\ell^* &=& \arg\max_{\phi_\ell \in [0,\frac{2\pi}{M})}  \min_{m_\ell} \hat{\lambda}_{m_\ell,\phi_\ell}\\
\lambda_{m_\ell}^* &=& \hat{\lambda}_{m_\ell,\phi_\ell^*} \\
\mbox{where,},\; \hat{\lambda}_{m_\ell,\phi_\ell}  &=&  \frac{\prod_{b\in S_{m_\ell,\phi_\ell}} \frac{T_b}{Cap(p_\ell,p_b,\phi_\ell,\phi_b)}}{\sum_{b\in S_{m_\ell,\phi_\ell}} \frac{\prod_{b\in S_{m_\ell,\phi_\ell}} \frac{T_b}{Cap(p_\ell,p_b,\phi_\ell,\phi_b)}}{\frac{T_b}{Cap(p_\ell,p_b,\phi_\ell,\phi_b)}}} 
\end{eqnarray}
\normalsize
In addition  to contraction within each sector of the GS UAV, \system also explores contraction across the edges of adjacent sectors (sep 37 in Alg.~\ref{alg:skyhaul}). It executes the contraction procedure again within each virtual sector, where the latter consists of two adjacent sectors. 

 \subsection{Performance Monitoring and Adaptation}
\system adapts its backhaul network configuration from time-to-time to
cater to the varying positions and traffic demands of application UAVs
that are tracking the target event.  It monitors the end-to-end throughput
of each session at the GS along with the energy levels of UAVs.
It triggers a network
configuration update, when (i) the current demand satisfaction of any session
drops below a certain threshold $Th_a$ (balances responsiveness vs. session interruptions), (ii) UAVs need energy replenishment (when energy drops below $Th_e$), or (iii) application UAVs have been reassigned to cover other areas (ROIs).  
\system first attempts to find a configuration with updates to only the yaws of the UAVs that carry traffic for the affected sessions (i.e., angular optimization without contraction),
thereby avoiding UAV migrations. 
 When this is not feasible, 
 it executes a complete configuration update that might necessitate UAV migrations. 


\subsection{Seamless UAV Migration Plan} 
\label{subsec:uav-migrate}

\setcounter{algorithm}{1}
\begin{algorithm}
    \small
    \caption{De-conflicted Configuration Migration}\label{alg:configmig}
    \begin{algorithmic}[1] 

\State Determine min-max bipartite matching assignment between old and new
configuration UAV positions (with Euclidian distance as weights), which
minimizes the time taken to migrate to the new configuration.

\State Color the conflict graph created with matching assignments as vertices
and crossing assignments (in the Euclidian plane) as conflicts (edges).

\State Assign configuration changes (edges) with different colors to different
altitudes to get a collision-free flight control path.

\State Move the UAVs (assigned to different altitudes) in altitude (z-axis) alone first, while retaining their position in the x-y plane; then let the UAVs in different altitudes move to their new position in the x-y plane as dictated by their new configuration; then bring them all back to the original altitude for operation in the new configuration.
\end{algorithmic}
\end{algorithm}
\normalsize

\begin{figure}
    \centering
    \includegraphics[width=0.7\columnwidth,height=3.5cm]{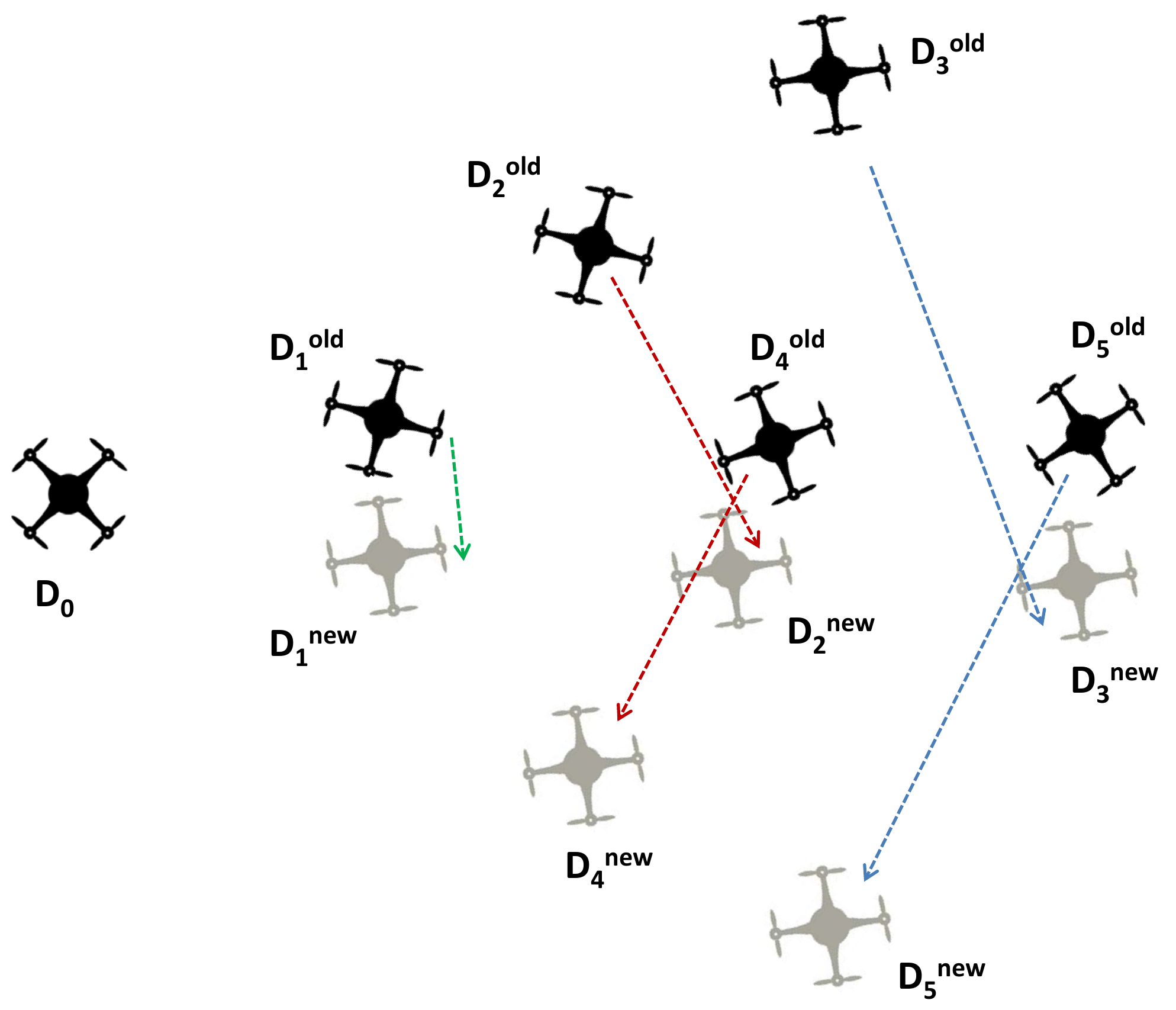}
    \vspace{-0.1cm}
    \caption{Adaptation of backhaul configurations.\label{fig:configmig}}
    \vspace{-0.6cm}
\end{figure}

Whenever a configuration update is triggered, \system computes a new configuration 
(Alg.~\ref{alg:skyhaul}) based on current position and
demands from application UAVs. However, the biggest challenge in realizing this
new configuration lies in UAVs optimally (i.e., in least time) migrating from their old to new
configurations (as shown in Fig.~\ref{fig:configmig})
without any collisions in the airspace. \system accomplishes
this migration through a novel, optimal de-confliction flight control algorithm
(Alg.~\ref{alg:configmig}) by leveraging {\em various layers of altitude} as
described below. This same approach can be used to handle deployment of UAVs
(relay and application) both during network initialization and later, as well as
recovery of existing UAVs (for battery replacement) as special
cases

{\bf Flight plan for minimum migration time} (Step 1): \system first determines
the flight plan that would result in the least duration required to migrate to
the new configuration, without accounting for potential conflicts between UAV
movements during the migration (to be resolved in the second step).  It
accomplishes this by formulating the flight plan problem as a variant of the
bipartite matching problem, where the objective is to determine the min-max
bipartite matching assignment between old and new configuration UAV {\em
positions} (with Euclidian distance as weights, serves as a proxy for migration
time).  The  min-max objective helps us pick the migration plan that  minimizes
the maximum of the weights, i.e., longest time taken by any of the UAVs during
their simultaneous migration.

\begin{figure}
    \centering 
    \begin{subfigure}{0.52\columnwidth}
        \centering
        \includegraphics[width=\columnwidth]{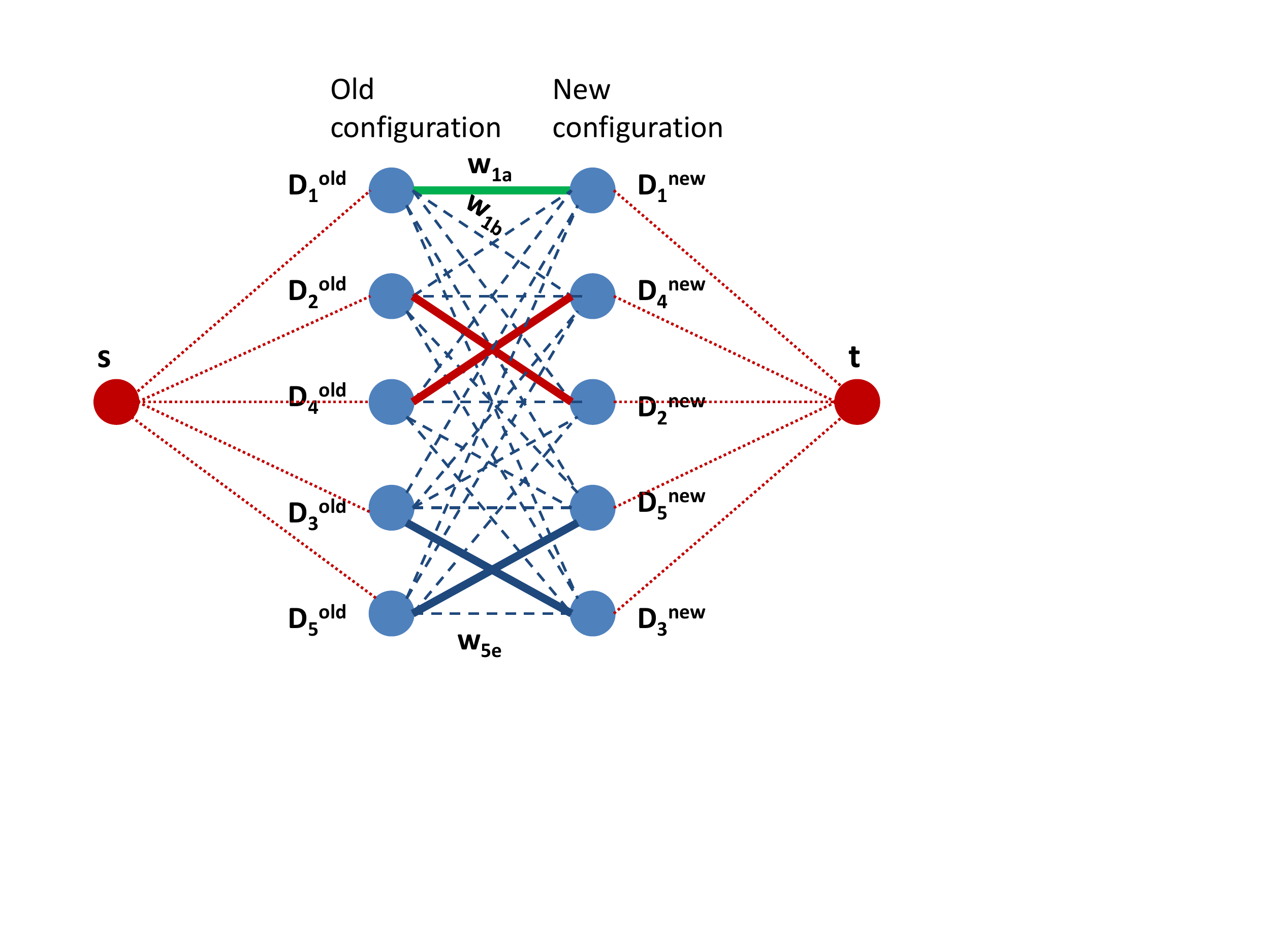}
        \caption{Min-max matching between old and new configurations.}
    \end{subfigure}
    \hspace{2pt}
    \begin{subfigure}{0.45\columnwidth}
        \centering
        \includegraphics[width=\columnwidth]{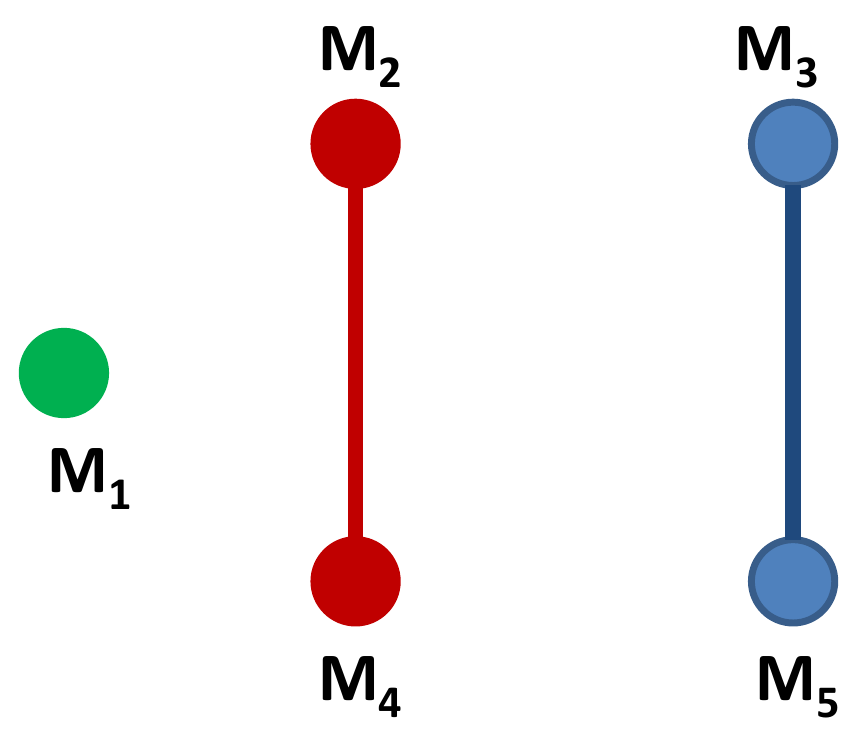}
        \caption{Matching/migration conflicts in Euclidian space.}
    \end{subfigure}
    \vspace{-3pt}
    \caption{Conflict-free migration plan.\label{fig:migplan}}
    \vspace{-19pt}
\end{figure}

\begin{figure*}
    \centering
    \begin{subfigure}[b]{0.65\columnwidth}
      \includegraphics[width=1.1\textwidth, height=3cm]{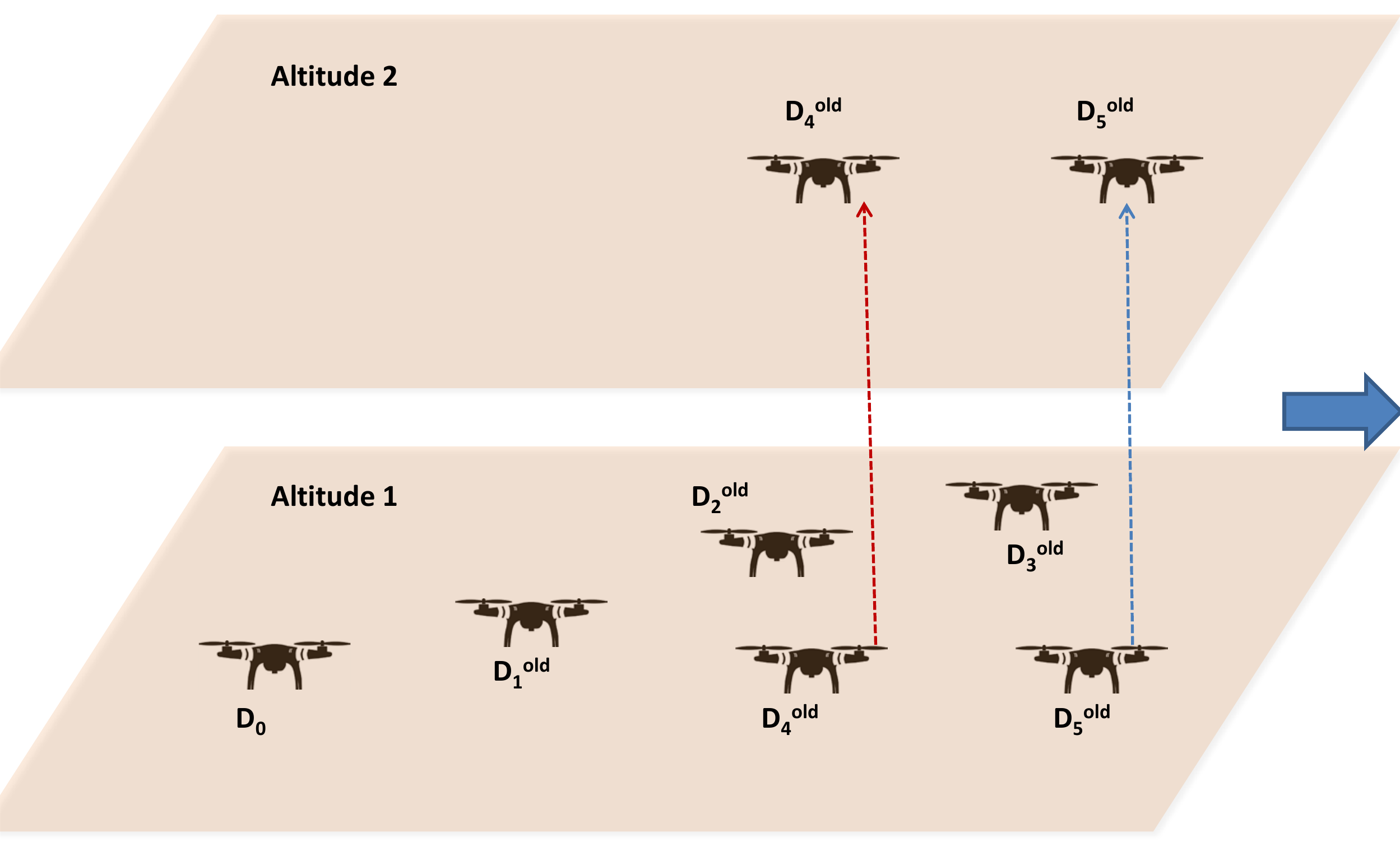}
    \caption {{Move conflicting migrations to different altitudes.}}
    \end{subfigure}
    \hspace{0.1cm}
    \begin{subfigure}[b]{0.65\columnwidth}
    \includegraphics[width=1.1\textwidth, height=3cm]{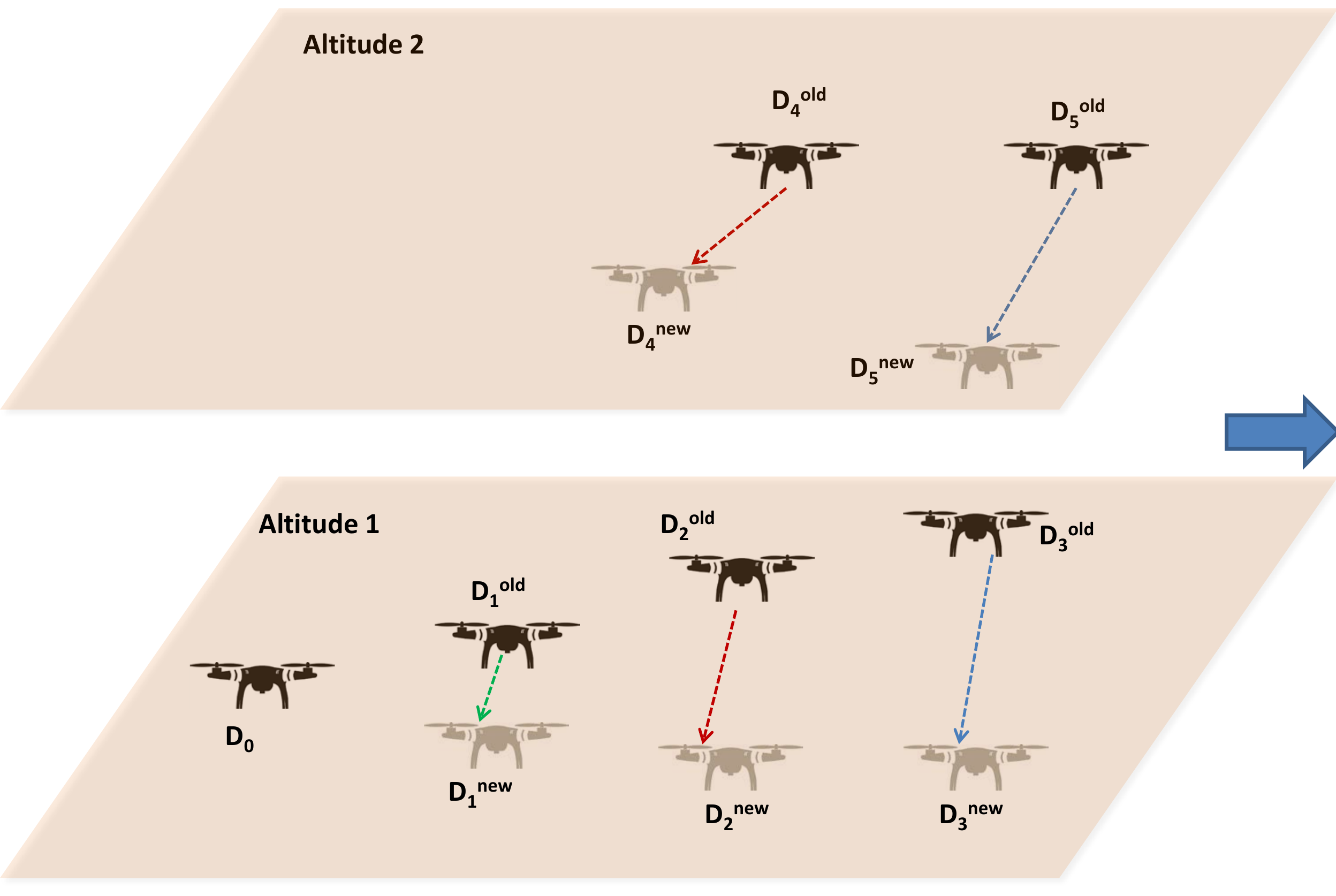}
    \caption{Move to new configuration in respective altitudes.}
    \end{subfigure}
        \hspace{0.1cm}
    \begin{subfigure}[b]{0.65\columnwidth}
    \includegraphics[width=1.1\textwidth, height=3cm]{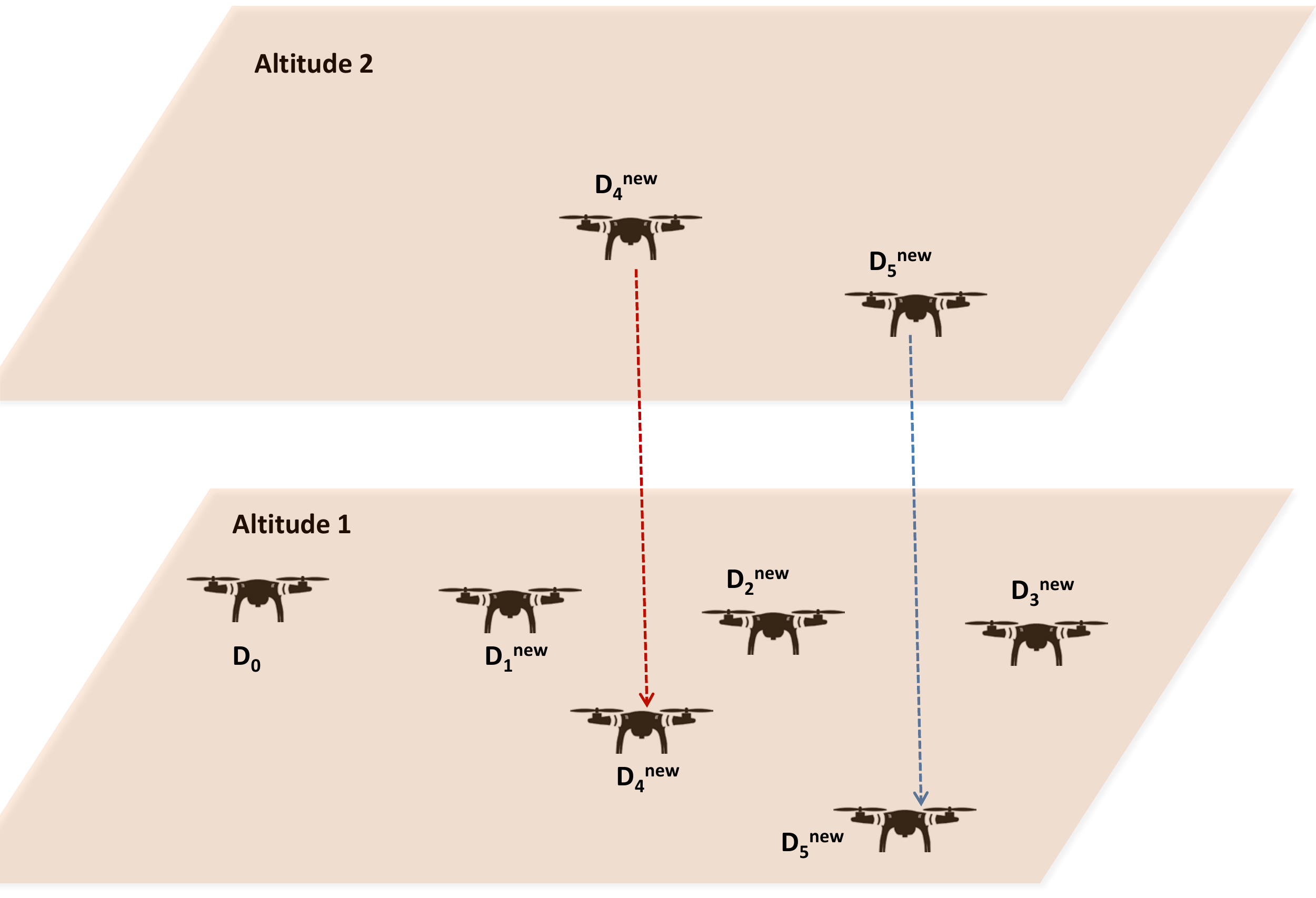}
    \caption{Lastly, bring back all UAVs to operational altitude.}
    \end{subfigure}
    \vspace{-0.4cm}
    \caption {{Conflict-free migration to new configuration.} \label{fig:migration}}
    \vspace{-12pt}
\end{figure*}

\underline{\em Migration algorithm:}  While algorithms exist for maximum or
minimum weight bipartite matching, the variant of min-max bipartite matching
(M2BM) has not been addressed in literature. To this end, we now provide an
optimal algorithm to solve M2BM.  The algorithm proceeds by forming a flow graph
$G''$ from the bipartite graph $G$, using the same set of vertices but adding a
source and sink that are connected to the vertices in $V_o$ and $V_n$
respectively. Then it sorts the edges in $G$ based on increasing weights and
adds them one by one to $G''$. At each iteration, a max-flow algorithm is run
between $s$ and $t$ on $G''$ with all edges carrying a weight of one. The
resulting max-flow solution yields a maximum matching $M$ between vertices $V_o$
and $V_n$ in $G$. If the matching is perfect in $G$, i.e., its size is equal to
that of the vertex set of $G$, then $M$ is the desired migration configuration
(see Fig.~\ref{fig:migplan}). Otherwise, the next sorted edge from $G$ is added
to $G''$ and the process is repeated until a perfect matching is found. The
resulting  solution is optimal 
in that it takes the least time for configuration
migration.

{\bf De-conflicting the flight plan} (Steps 2-4): The migration solution does
not account for potential conflicts/collisions in the airspace when UAVs
migrate in tandem on the same Euclidian plane (i.e., when their migration paths
cross).  To deconflict such colliding migration paths, \system uses the
layers of altitude available for migration (steps 2-4,
Alg.~\ref{alg:configmig}).  Specifically, \system creates a conflict graph
$G_c$, with the edges of the output matching $M$, now serving as vertices
($V_c$) of $G_c$, while edges in $G_c$ correspond to migration paths ($V_c$)
that conflict in the Euclidian plane (e.g. paths that cross or come in close
proximity of one another). \system then colors $G_c$ with the least possible
colors (step 2, Alg.~\ref{alg:configmig}), and  assigns migrations (vertices)
with different colors to different altitudes to get a collision-free flight
migration path (step 3, Alg.~\ref{alg:configmig}). Then, it executes the
migration in a three step procedure as shown in Fig.~\ref{fig:migration} (step
4, Alg.~\ref{alg:configmig}): (i) move the UAVs (assigned to different
altitudes) in altitude (z-axis) alone first, while retaining their position in
the x-y plane; (ii) Then, allow the UAVs in different altitudes to move
to their new position in the x-y plane as dictated by their new configuration;
and (iii) finally, bring them all back to the original altitude for operation
in the new configuration.

In practice, owing to Euclidian constraints and a migration solution targeting
minimum duration, the conflict graph $G_c$ hardly has cliques of size three or
more. This allows for at most three altitude layers (existing, one above, one
below) needed to resolve the conflicts.  Thus, the bulk of the time needed for
configuration migration arises from moving to the new configuration at a
prescribed altitude (time to move across altitude layers is relatively
insignificant), which in turn is optimized by \system.

{\bf Impact of Altitude Changes}: \system operates with UAVs on the same
\emph{operating altitude}.  During UAV migration, \system may temporarily move
the UAVs to different altitudes away from the \system operating
altitude. This will temporarily disconnect the UAV from the
\system mesh, and will reconnect upon UAV returning to the operating altitude. 
These temporary disconnections are short and do not impact the overall operation.

\subsection{Performance and Time Complexity}
{\bf Time Complexity:} We first characterize the time complexity of \system's configuration determination algorithm, Alg.~\ref{alg:skyhaul}.
The complexity in the radial optimization comes from solving the convex optimization for initial deployment per session, which is $O(K^3)$, followed by orientation determination at each of the drones,    which is $O(N)$. Here, $K$ and $N$ are the number of sessions and drones per sector of the GS drone.
Given that this optimization is done in each of the GS drone's (logical) sectors (total of $M$) independently, this results in a net complexity of $O(M\cdot (K^3+N))$.
The key time consuming component in angular optimization arises from the contraction step, whose worst case complexity results when every pair of drones (belonging to different sessions) needs to be      considered for contraction without success. This results in a complexity of $O(N^2)$, while the check of feasible orientation contributes another $O(N)$ per pair considered, resulting in a net complexity of $O(M\cdot N^3)$ for angular optimization. Putting the two steps together, \system accomplishes network configuration determination with an efficient complexity of $O(M\cdot (K^3+N^3))$, where the      cubic impact is only with respect to sessions and drones in a sector and not the whole network.

The time complexity of \system's configuration migration algorithm, Alg.~\ref{alg:configmig} is dominated by the computation of the migration plan itself. This in turn requires computing a max-flow       solution for the network graph (with $O(N)$ vertices, where $N=|V|$), whose edges are added incrementally.
The complexity of the max-flow solution is $O(|E|\cdot f)$, where $E$ is the number of edges (that varies incrementally from $O(N)$ to $O(N^2)$) and $f$ is the max-flow value, which in our case is the    size of the perfect matching ($N$). Hence, when accounting for all the sectors, this results in a net complexity of $O(M\cdot (N^4))$. We believe this can be reduced to $O(M\cdot N^3)$ by incrementally   growing the augmenting path of the max-flow solution in each iteration without having to run the whole max-flow algorithm.

{\bf Performance:}
Given the hardness of the configuration computation problem, it is hard to establish a performance guarantee for Alg.~\ref{alg:skyhaul}.
However, as we show in the evaluation section, \system delivers very efficient performance in practice -- it is able to satisfy all the traffic demands of the target application at a deployment cost that is very close to that of a (minimum) Steiner point solution; the latter serving as a bound on the minimum number of drones needed to be deployed purely from a binary connectivity standpoint without       accounting for mmWave constraints or traffic demands.

With respect to \system's migration solution, we can establish the following performance.

\system determines an optimal migration plan in $O(M\cdot N^4)$, where $M$ is the number of logical sectors (radios) at the GS drone, and $N$ is the number of drones deployed per sector.
\begin{proof}

Having already shown the time complexity, this reduces to showing that Alg.~\ref{alg:configmig} computes a perfect matching (of size $N$), whose maximum weight (migration time) among all its migrations   is the minimum. By incrementally introducing edges,  \system ensures that a maximum matching is found at every iteration. Hence, if there exists a perfect matching at any iteration, it will be found.     Given the addition of edges in order of increasing weights, this automatically results in the smallest (max) weight for which a perfect matching is possible.
\end{proof}

\subsection{Leveraging Multiple Ground Stations}
\label{sec:ext}

\system can be easily adapted to leverage multiple ground stations.
Note that while multiple GSs bring additional capacity (entry/exit points) to the backhaul, they would also incur additional drone deployment for connectivity. The challenge is to leverage the            appropriate number of GSs (and their capacity) to support the desired traffic demands of the application.

Given a set of potential GSs (and their ground locations), \system accomplishes this by incrementally picking one GS in each iteration, where the GS that yields the highest marginal utility               ($\frac{\min_{k\in \mathcal{K}}\lambda_k}{|\mathcal{D}_{{cur}}|}$) is chosen (from remaining GSs) and added to the network.  GSs are added until desired traffic demands can be satisfied by the network.

In each iteration described above, \system needs to solve the backhaul deployment problem for a given number of GSs.
Here, the traffic demands to/from the application drones are equally split to the different GSs. Then, \system finds the initial  deployment of drones (using its radial  optimization) with respect to     each of the GSs in isolation. Then, it aggregates the deployment across all the GSs and contracts the backhaul drones (using its angular optimization) across both sessions and GSs jointly. This allows    \system to determine an efficient backhaul deployment that is tailored to the positions of the GSs under consideration.


%% file: evaluation3.tex
We implement \system on four DJI Matrice 600 Pro UAVs to
demonstrate its performance in realistic flight conditions, and
highlight its ability for autonomous reconfiguration of a multi-UAV
network. To understand the merits of \system's algorithms, we also study its
performance in large scale topologies (more UAVs) through simulations in
\S\ref{sec:simulate}. Video demonstration of \system can be found
at~\cite{videodemo}.

\subsection{Implementation}
\label{sec:impl}

\textbf{AWS.} \system controller is executed on an Amazon Web Services instance, and
communicates with the UAVs via an LTE interface (through a smartphone). Each UAV
continuously transmits all its telemetry data (i.e. GPS position, yaw, battery
level etc) to the AWS instance and executes movement instructions that are sent
back to them from \system controller.
However, when cellular services are unavailable, the 60Ghz network itself can 
be used as a control channel, with the \system controller running on the Ground
Station (GS) computer. The large bandwidth of the 60Ghz link ensures negligible overhead for
communicating control information along with the actual data.

\begin{figure}
    \centering
    \includegraphics[width=0.8\columnwidth]{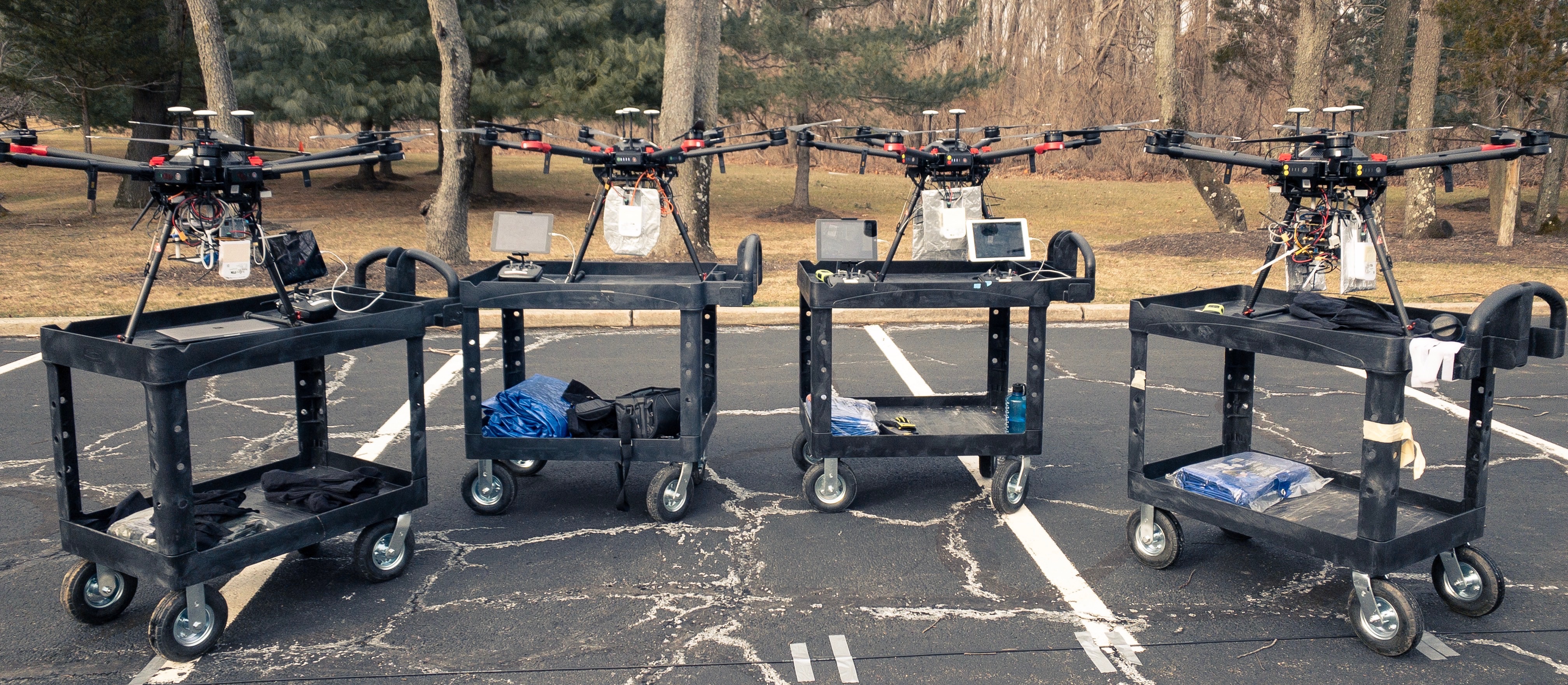}
	\caption {Picture of 2 Application, Relay \& GS UAVs \label{fig:UAV_pic}}
\vspace{-0.6cm}
\end{figure}

\textbf{M600Pro.} Each UAV is equipped with a single-board computer (SBC) with an
Intel Core i7-6600U CPU running Ubuntu 18.04. The UAV is 
controlled from this SBC using DJI OnBoard SDK. We design and install role-specific
payloads on the UAV platforms, one GS UAV, one relay UAV and two application
UAVs (Fig.~\ref{fig:UAV_pic}). The GS UAV has three 60GHz radios, one
pointing downwards towards the GS, and two mounted horizontally to
connect to the other UAVs. The relay UAV has two horizontally mounted radios
and application UAVs have one radio and a 4K camera for live video-streaming.  
Each 60GHz radio is connected to the SBC via an ethernet port. 


\textbf{UAV power consumption:} Each UAV has a flight time of
$\approx$20 min and can carry a payload 
of max. 5Kg. Power consumed by on-board 60Ghz radios is negligible compared to the
power required for the actual flight. Many 
long-endurance UAVs~\cite{uav1},\cite{uav2},\cite{uav3} with high-capacity batteries and 
flight times extending to many hours are already available for use. 

\subsection{\textbf{\system -- Evaluation} \label{sec:eval}}

We first evaluate the impact of mmWave link throughput due to UAV mobility.
Subsequently, we focus on evaluating \system, by running a series of experiments that show workings of the proposed algorithms
under two realistic scenarios that enable (i) on-demand communication (wireless access), and (ii) sensing (live 4K video analytics), from application UAVs.
We demonstrate \system's ability to seamlessly maintain network performance under actual
flight operations: The application UAVs move along a pre-determined path 
to emulate tracking of application dynamics (e.g. user movement, spread of forest fire),
while \system
continuously and autonomously positions the relay UAV to satisfy the high
traffic demands of the application UAVs.
We compare \system to a \textsc{MinDrone} baseline that 
only adjusts the yaw of the GS UAV to best maintain
connectivity and share its multiple radios with the application UAVs, without using a relay UAV.
This is analogous to most commercial solutions that currently exists.
When the application UAVs start moving,
\textsc{MinDrone} is unable to maintain the initial
level of network throughput with either the application or the GS UAV.

\begin{figure}
    \centering
    \begin{subfigure}[b]{0.47\columnwidth}
        \includegraphics[width=\textwidth,height=2.8cm]{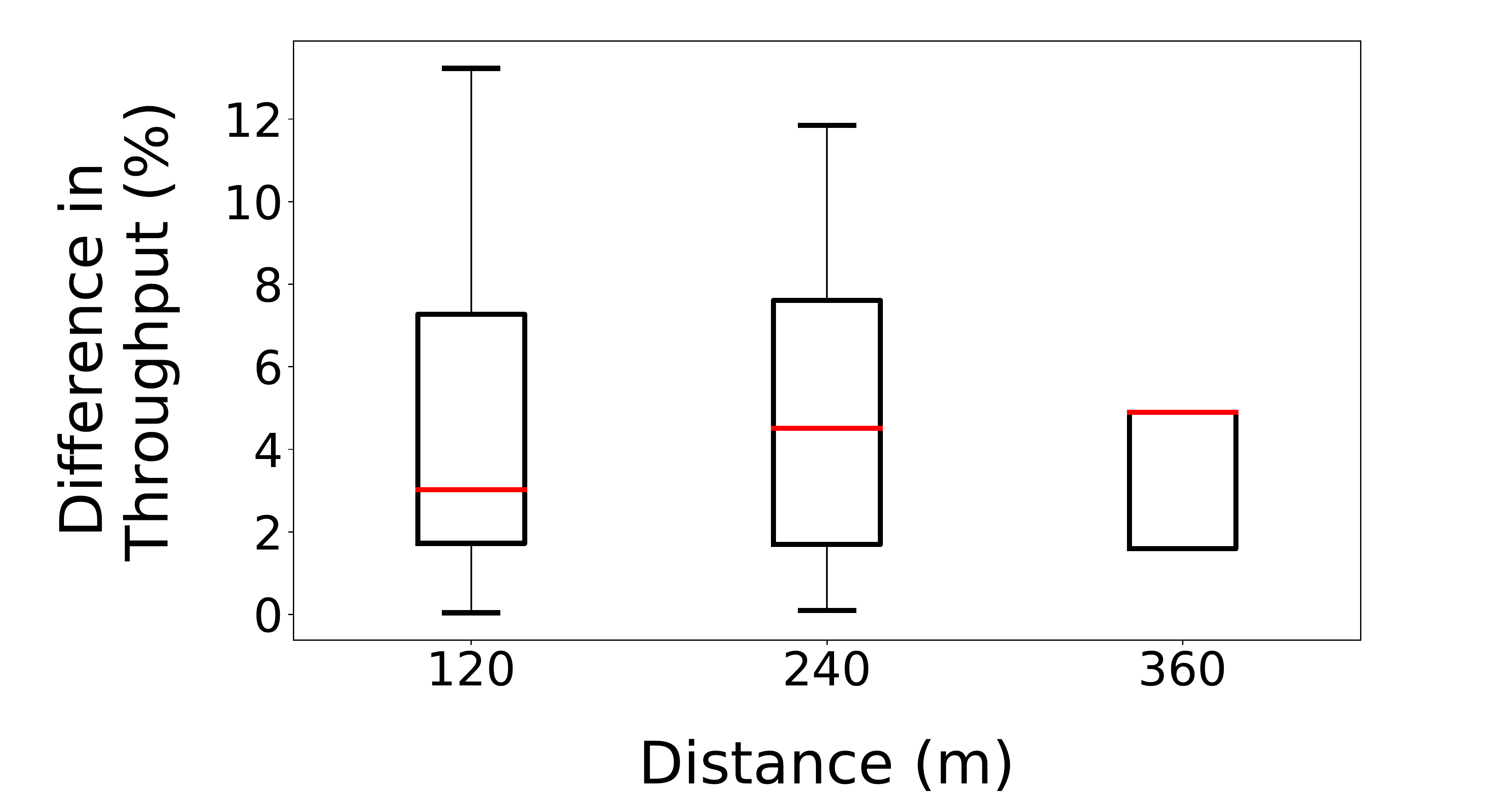}
    \end{subfigure}
    \hspace{2pt}
    \begin{subfigure}[b]{0.47\columnwidth}
        \includegraphics[width=\textwidth,height=2.8cm]{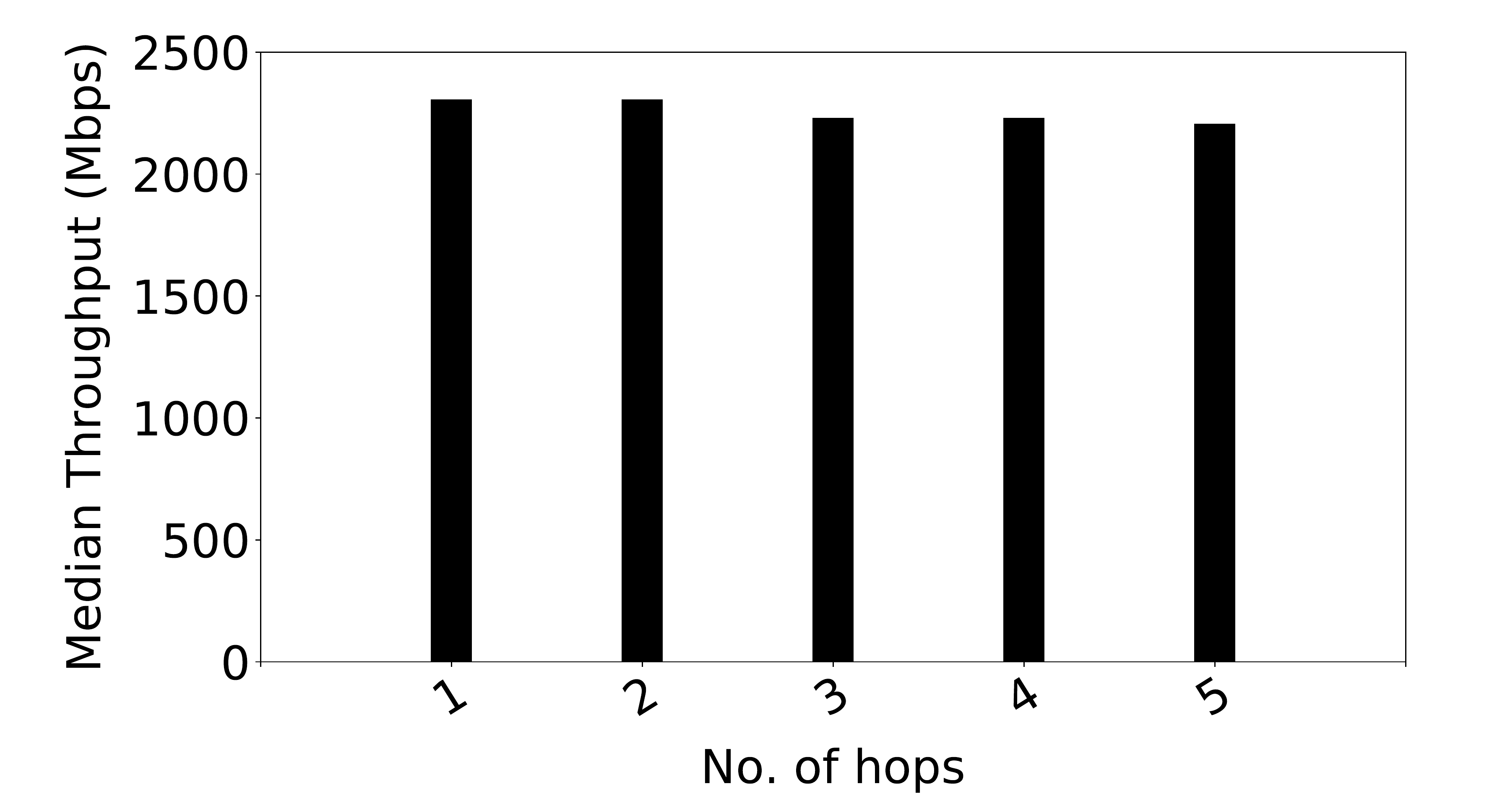}
    \end{subfigure}
    \vspace{-0.3cm}
    \caption { {Static Vs. Mobile--Difference in median throughput for all UAV Yaw (left) and Throughput Vs. Hop count (right), 
    with difference in radio angle between sender and receiver equal to \ang{0}.} 
    \label{fig:micro_exp}}
    \vspace{-0.5cm}
\end{figure}

\emph{\bf{Link Performance vs. UAV mobility:}}
\label{sec:link_mob}
The graph in Fig.~\ref{fig:micro_exp} shows the difference
in a link's median UDP throughput when two UAVs are static (hovering), and when they are continuously mobile
while maintaining a constant distance between them. The results are obtained for distances of
120m, 240m and 360m and for all
relative UAV orientations (yaws) from $\ang{-80}$ to $\ang{80}$ (in steps of $\ang{20}$) for each of the distances.
We do not observe a significant impact of UAV mobility on link throughput. This indicates that once the radios are oriented appropriately, 
small UAV vibrations and movements due to air turbulence
can be handled by 802.11ad's underlying beam selection algorithm. Additionally, we observe that the
end-to-end throughput is not affected by the path-length (\# of hops). Hence, extending coverage by adding multiple relay UAVs
should not affect \system's performance appreciably.


\begin{figure}
    \centering
    \begin{subfigure}[b]{0.47\columnwidth}
        \includegraphics[width=\textwidth,height=3.5cm]{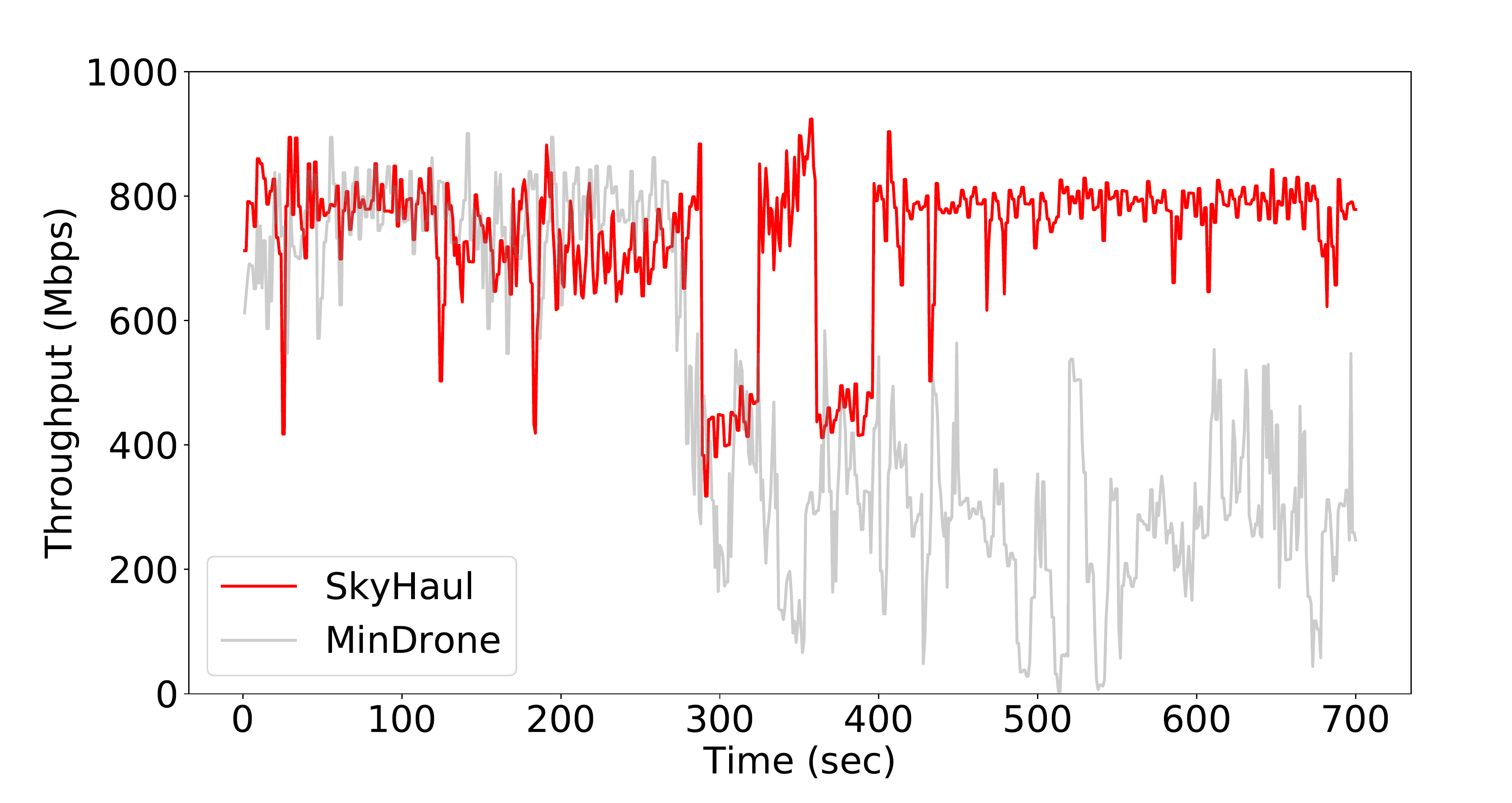}
    \end{subfigure}
    \hspace{2pt}
    \begin{subfigure}[b]{0.47\columnwidth}
        \includegraphics[width=\textwidth,height=3.5cm]{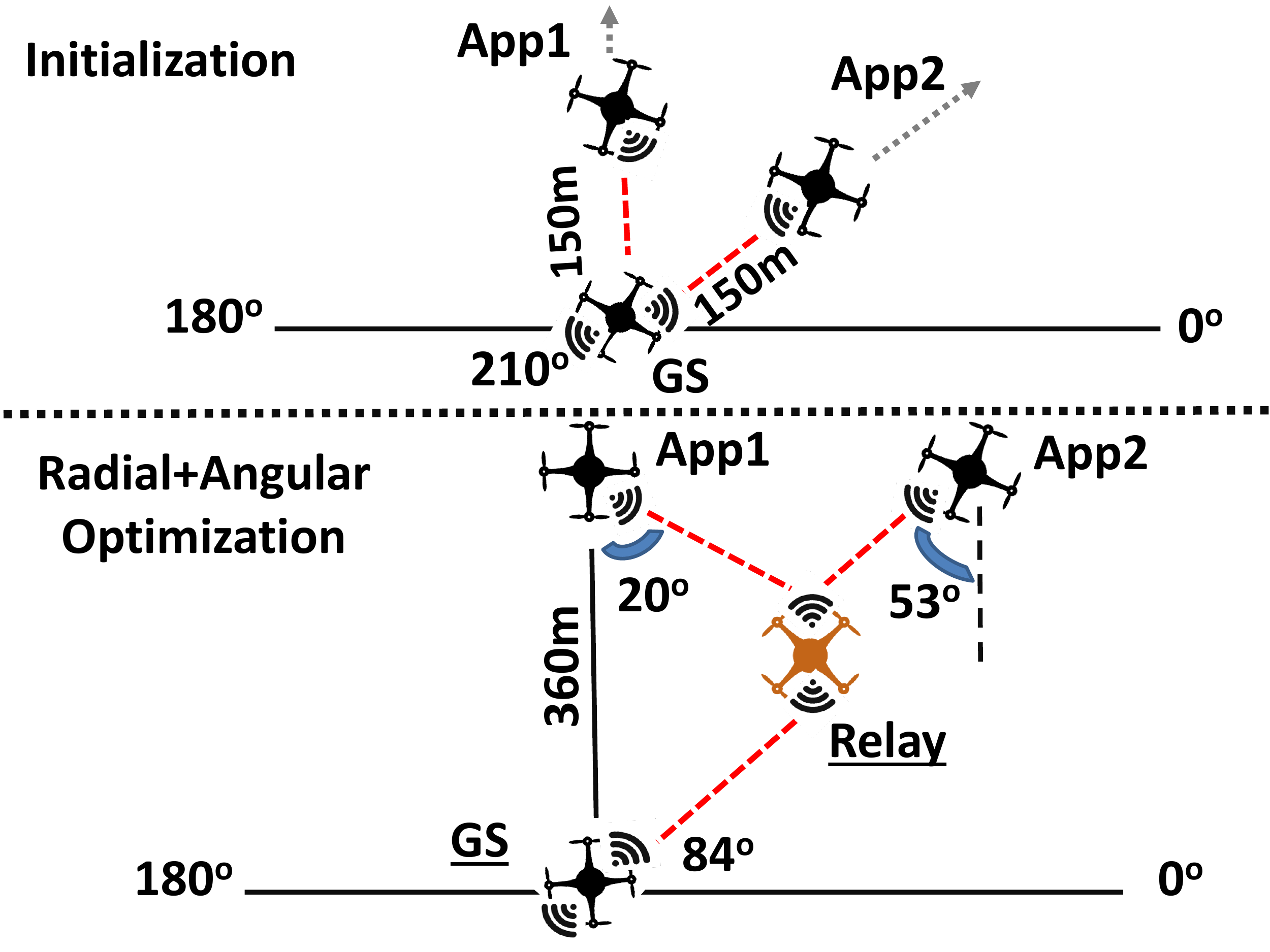}
    \end{subfigure}
    \vspace{-0.3cm}
    \caption { {\system Radial+Angular optimization - Throughput performance(left) and UAV placements (right).} 
    \label{fig:seg1}}
    \vspace{-0.5cm}
\end{figure}


\subsubsection{On-demand Wireless Access}
\label{sec:xput_eval} 
In order to emulate traffic load from a mobile small-cell (WiFi-AP/LTE),
each application UAV generates
450Mbps UDP traffic using \texttt{iperf3}. We
measure the actual received throughput at the GS. All the UAVs  operate at
60m altitude. 
\textbf{\system - Radial + Angular Optimization:} In our first experiment, we
initially place the application UAVs (\texttt{APP1} and
\texttt{APP2}) at \SI{150}{\metre} radius and \ang{90} and \ang{78} respectively from the GS UAV (fig.\ref{fig:seg1}).
With \texttt{APP1} and \texttt{APP2} capable of direct connectivity to the GS UAV, \system
determines the optimal yaw for the GS UAV (\ang{30}) and application UAVs (\ang{300}, and \ang{258}) that allows  
net traffic demand from both the application UAVs to be satisfied by its same 60GHz radio.
Since a relay UAV is not required 
yet, \system and  \texttt{MinDrone}
achieve a similar performance with an aggregate throughput of 800Mbps.

After couple of minutes, the application UAVs while maintaining their
angular offsets (w.r.t. the GS UAV), start moving until they are
\SI{360}{\meter} away from the GS UAV. Once the throughput of
\texttt{APP1} and/or \texttt{APP2} falls below a certain threshold
($\lambda=0.75$, 330Mbps), \system runs a configuration update through its
radial+angular optimization to determine that while relay UAVs are needed to
sustain service to either of the application UAVs, this can be achieved with a
single relay UAV. As shown in Fig.~\ref{fig:seg1}, it launches a relay UAV and
positions it at \ang{84} and 160m from the GS UAV.  The orientation of the relay
and application UAVs are then updated to \ang{90}, \ang{290} (\texttt{APP1}),
\ang{217} (\texttt{APP2}), respectively, such that one of the relay UAV's radios
is used to jointly serve both the application UAVs, while the other provides
connectivity to the GS UAV, latter's yaw being \ang{84}.  During this dynamic
event, we observe a sharp throughput degradation in Fig.~\ref{fig:seg1}
that results from disconnection as well as re-connection through
the relay UAV.  However, after \system completes its reconfiguration, the
aggregate throughput is immediately restored to 800Mbps.  Without the aid of a
relay UAV, \textsc{MinDrone} is unable to sustain the application demands,
providing an aggregate throughput of at most 400Mbps, while reaching a low of
200Mbps.

\begin{figure}
    \centering
    \begin{subfigure}[b]{0.47\columnwidth}
        \includegraphics[width=\textwidth,height=3.5cm]{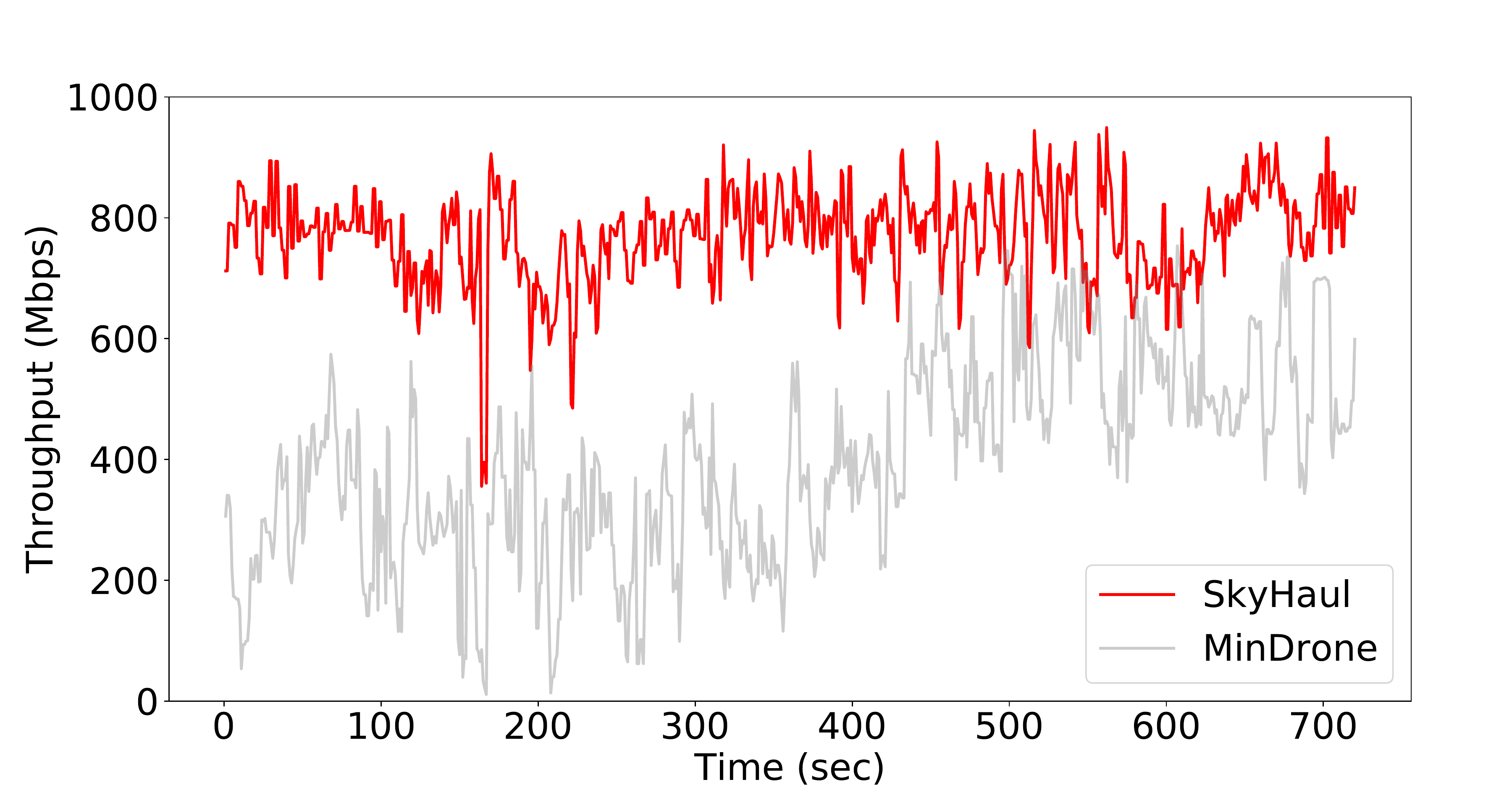}
    \end{subfigure}
    \hspace{2pt}
    \begin{subfigure}[b]{0.47\columnwidth}
        \includegraphics[width=\textwidth,height=3.5cm]{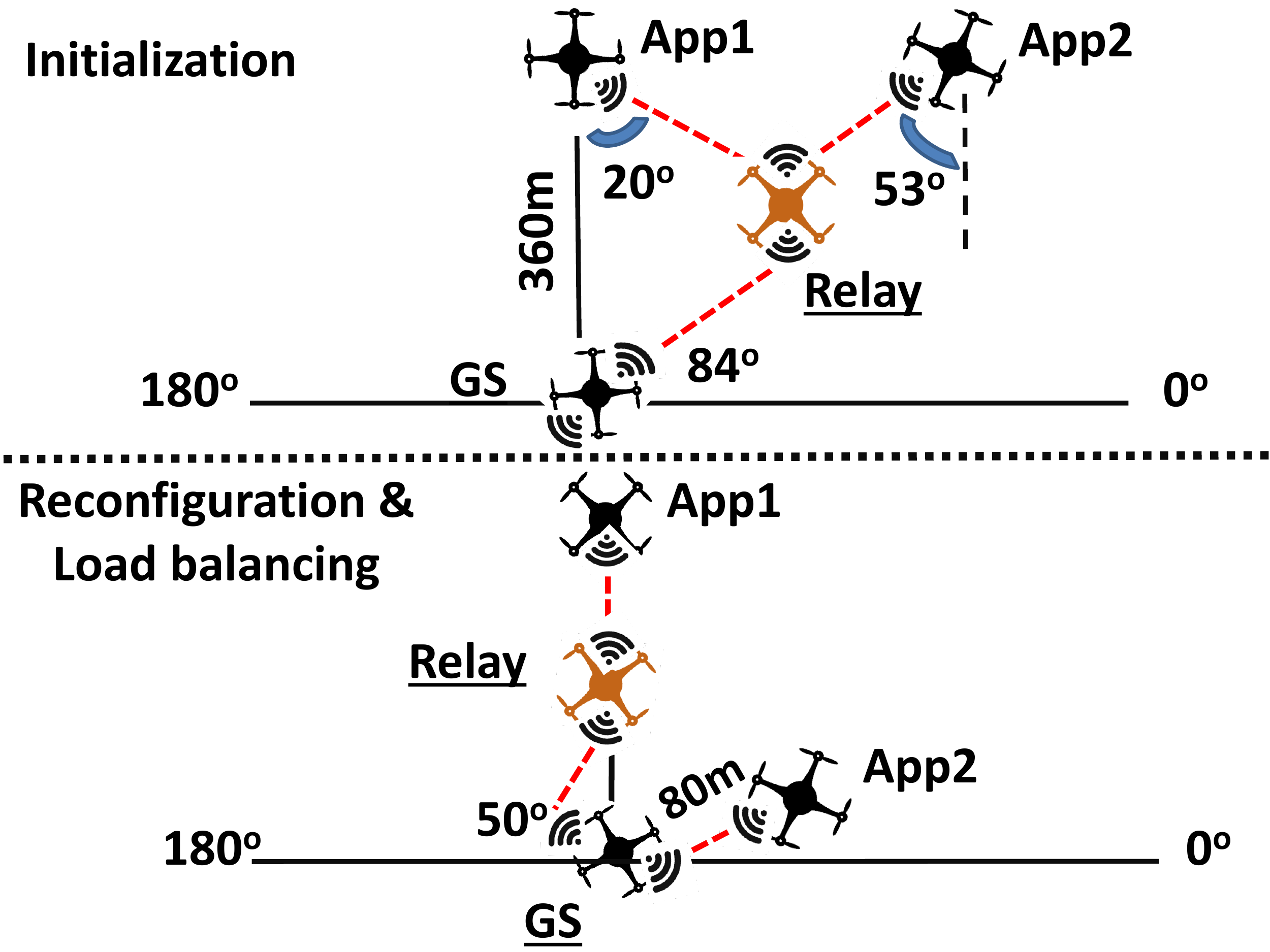}
    \end{subfigure}
    \vspace{-0.3cm}
    \caption { {\system Re-configuration and load balancing - Throughput performance (left), UAV placements (right).} 
    \label{fig:seg2}}
    \vspace{-0.5cm}
\end{figure}

\textbf{\system - Re-configuration and Load Balancing:} In the next experiment,
we initially place both \texttt{APP1} and \texttt{APP2} 360m away from the 
GS UAV, with both being served by a relay UAV placed as shown in the Fig~\ref{fig:seg2}.
While \texttt{APP1}
remains static, \texttt{APP2} starts moving closer to the GS UAV until it is
\SI{80}{\metre} apart (refer to Fig.~\ref{fig:seg2}).  \system determines
that while GS UAV can directly connect and support \texttt{APP2}'s traffic
demands,  \texttt{APP1} still needs a relay UAV.  It computes the updated
configuration,  resulting in relay UAV moving to 165m, at \ang{90} from GS
UAV to serve \texttt{APP1} alone, while the GS UAV updates its yaw to \ang{30}
allowing it to use two of its radios, each serving one 
application UAV. The application and relay UAVs update 
yaws to \ang{270},\ang{210} and \ang{270}, respectively.  This shows \system's
ability to efficiently leverage radios from various UAVs in the network to best
address application requirements.      

\begin{figure}
    \centering
    \begin{subfigure}[b]{0.47\columnwidth}
        \includegraphics[width=\textwidth,height=3.5cm]{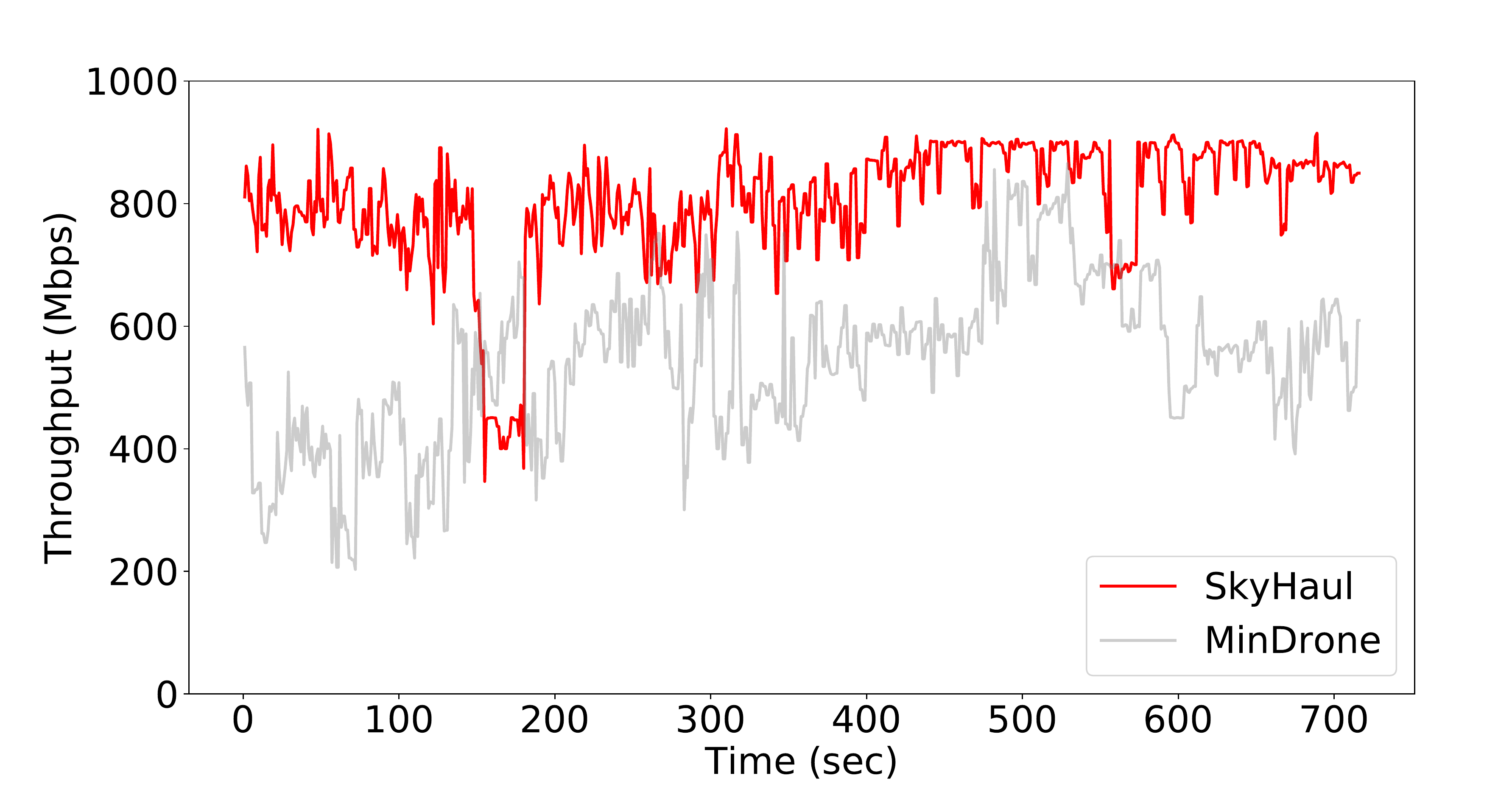}
    \end{subfigure}
    \hspace{2pt}
    \begin{subfigure}[b]{0.47\columnwidth}
        \includegraphics[width=\textwidth,height=3.5cm]{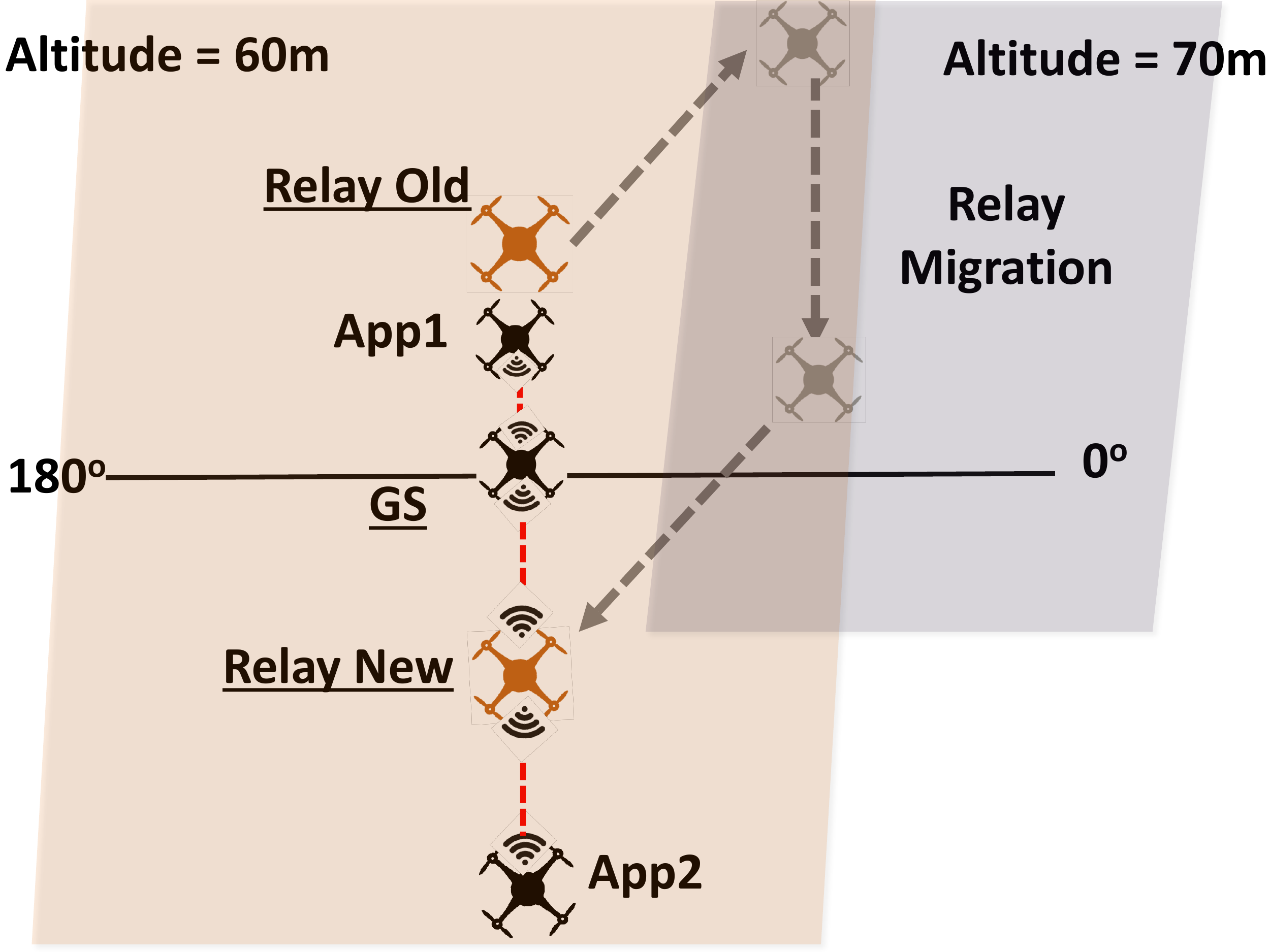}
    \end{subfigure}
    \vspace{-0.3cm}
    \caption { {\system Network migration - Throughput performance (left), UAV placements (right).} 
    \label{fig:seg3}}
    \vspace{-0.3cm}
\end{figure}

\textbf{\system - Configuration Migration:} In our third experiment, we initialize
\texttt{APP1} and \texttt{APP2} to be stationed at distances of 360m ($\ang{90}$) 
and 100m ($\ang{270}$) from the GS UAV, respectively. \system's initial configuration
requires \texttt{App1} to be served via a relay UAV and \texttt{App2} can connect
directly to the GS UAV. After a minute,
we simultaneously move \texttt{APP2} away from the GS UAV to
360m,\ang{270},
and  \texttt{APP1} closer towards
the GS UAV to a distance of \SI{100}{\metre}.   
With \texttt{APP1} capable of direct connectivity to the GS UAV, \system determines that the relay UAV needs to be re-purposed for \texttt{APP2}. 
However, this requires a 
flight plan, since the path of \texttt{APP1} conflicts with that of the relay UAV, and the latter's with the GS UAV.  
\system employs its efficient route-planning solution
to determine an optimal (fast) collision-free migration of the relay UAV towards \texttt{APP2}: the relay UAV first ascends to a
different altitude of 70m (10m above operating altitude of 60m) to resolve conflicts; while the relay UAV moves to its new position at
70m, the two application UAVs move to their new positions at 60m simultaneously; then the relay UAV reduces its altitude back
to operational height of \SI{60}{\metre}, while retaining its position in the plane. The whole migration process is executed 
autonomously with no human input. At this point, \system updates the yaw of
\texttt{APP2}, relay and GS UAVs to \ang{90} such that both application UAVs are effectively served.
The throughput result in Fig~\ref{fig:seg3} captures this transition, as the net-throughput decreases before
being restored after re-connection. \textsc{MinDrone} is unable to accomplish
such an adaptation and thus, suffers a drastic drop in throughput to about 500Mbps over
the same duration.

%% file: video.tex
\vspace{-0.4cm}
\subsubsection{Cloud-Based 4K Video Analytics} 
\label{sec:video}

We demonstrate \system's capability in aerial video surveillance applications by
conducting real-time video analytics on multiple 4K 60FPS video streams sent
from application UAVs to the GS UAV. Application UAVs follow the same
flight path as in \S\ref{sec:xput_eval}, while each sends a video stream using
\texttt{FFMPEG} to the GS, which are then
forwarded to our local GPU running YOLO3~\cite{yolo3} object detection algorithm.

\begin{figure}
    \centering
    \begin{subfigure}[b]{0.47\columnwidth}
    \includegraphics[width=1.1\textwidth]{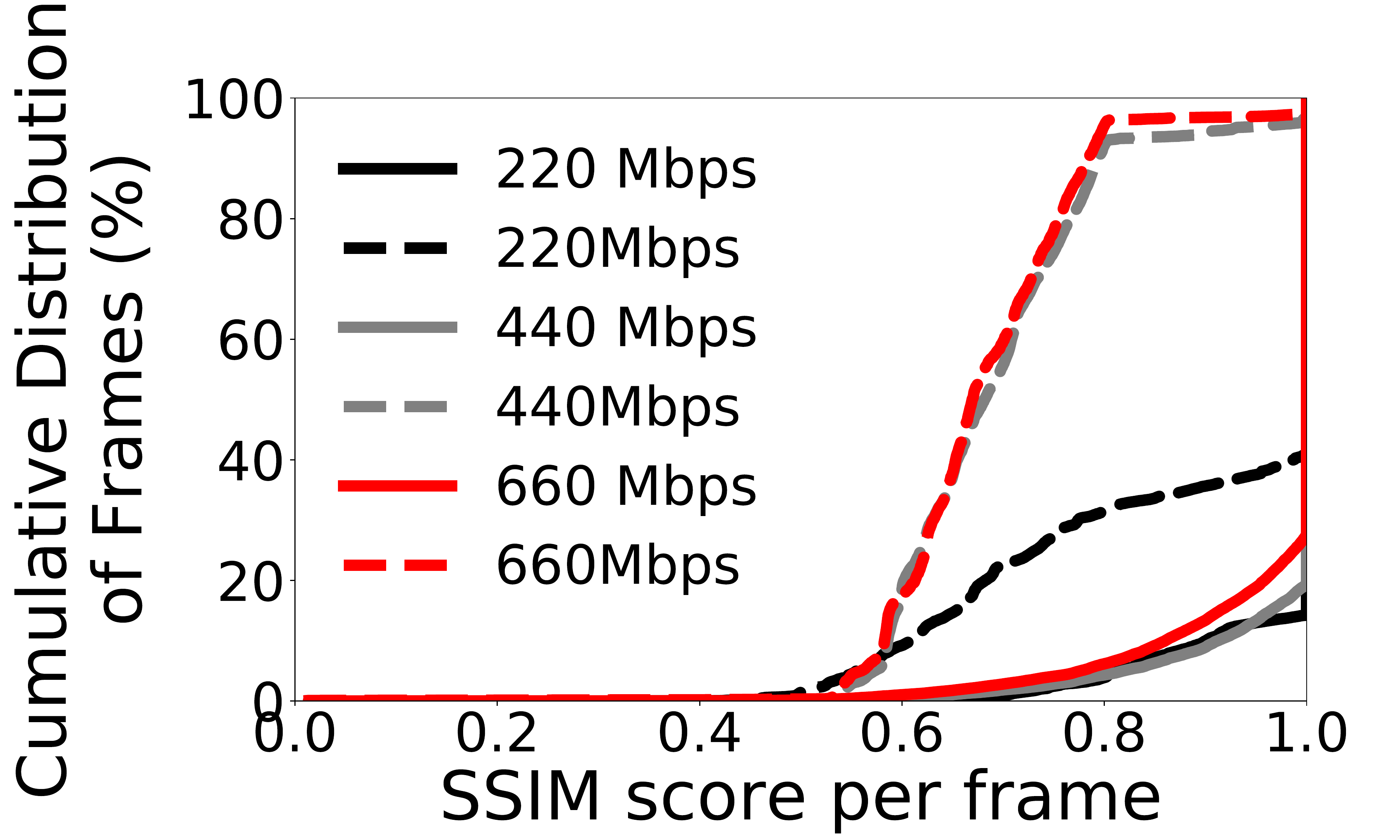}
	    \caption {\small {Per-frame SSIM index for \system (solid lines) and 
                \textsc{MinDrone} (dashed lines)} \label{fig:ssim_cdf}}
    \end{subfigure}
    \hspace{0.05cm}
    \begin{subfigure}[b]{0.47\columnwidth}
      \includegraphics[width=1.1\textwidth]{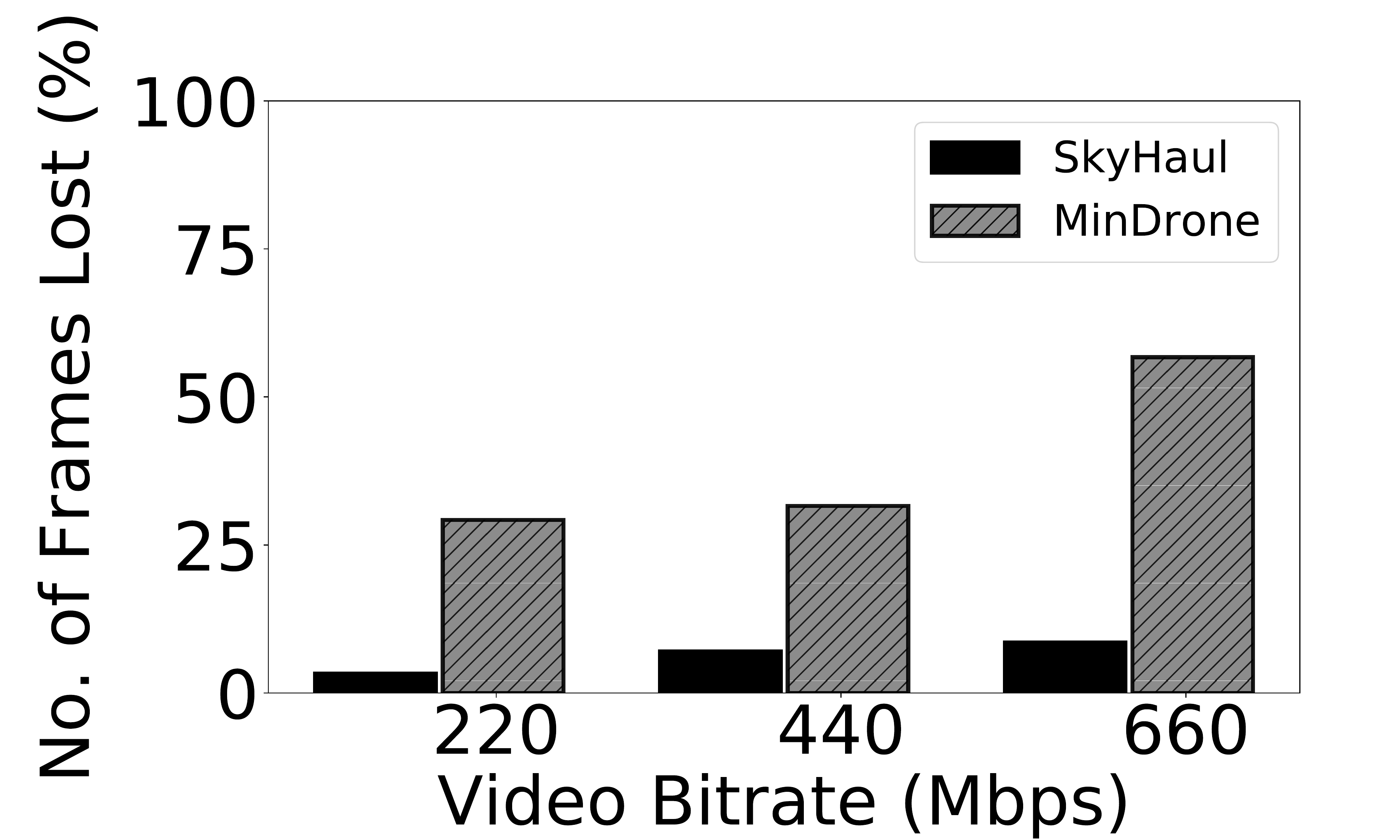}
    \caption {\small {Total No. of video frames lost in Baseline \& \system}  \label{fig:frames_lost}}
    \end{subfigure}
	\vspace{-0.3cm}
	\caption {{Video quality for \textsc{MinDrone} \& \system} \label{fig:eval_video}}
	\vspace{-0.5cm}
\end{figure}

\textbf{Received Video Quality.} Fig.~\ref{fig:ssim_cdf} compares the per-frame
Structural SIMilarity Index (SSIM)~\cite{ssim} of the video stream, encoded at
different  bitrates, received using \system and \texttt{MinDrone} w.r.t. the
original stream recorded locally at each application UAV. Compared to traditional metrics -- 
PSNR and MSE, SSIM metric takes into account more features between the original
and the received images for quality comparison. SSIM
considers three things: (a) Similarity in luminance, (b) Similarity in contrast, and
(c) Correlation between the two images.  The SSIM index is a value between 0 (no
similarity) and 1 (100\% similar) --- the higher the SSIM value, the ``cleaner''
the received image and thus, the better the object detection accuracy. 

Observe that \system achieves significantly higher SSIM values than
\texttt{MinDrone} at all video bitrates. While \system achieves an SSIM index of
1 for over 80\% of the frames with a bitrate of 660 Mbps,
\texttt{MinDrone} suffers early even at 220 Mbps bitrate, with only
15\% of frames achieving an SSIM score of over 0.8. As seen in Fig.~\ref{fig:frames_lost}, higher bandwidth of the
backhaul also helps to significantly reduce the number of lost frames in \system, 
which directly impacts on improving the accuracy of the object detection task.

\textbf{Object Detection.} 
We use a cloud server instance with a 32 core Intel E5-2686v4 CPU and
four NVIDIA Tesla V100 GPUs with \system for real-time video analytics. Video
streams from the application UAV are sent to the cloud instance from the
ground station for object detection.

\begin{figure}
    \centering
    \begin{subfigure}[b]{0.47\columnwidth}
        \includegraphics[width=\textwidth,height=3cm]{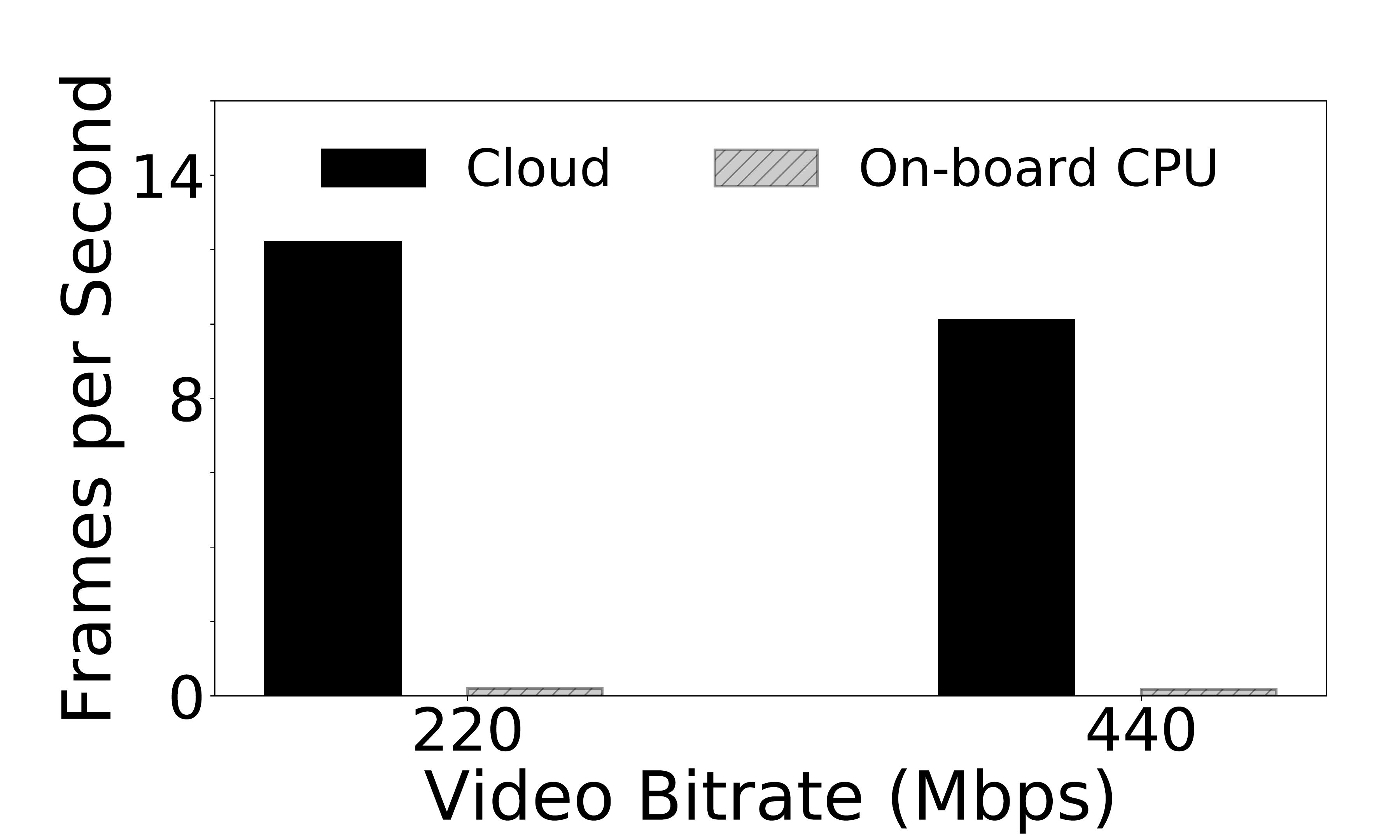}
        \caption {\small {Video processing speed in cloud and on-UAV CPU} \label{fig:fps}}
    \end{subfigure}
    \hspace{2pt}
    \begin{subfigure}[b]{0.47\columnwidth}
        \includegraphics[width=\textwidth,height=3cm]{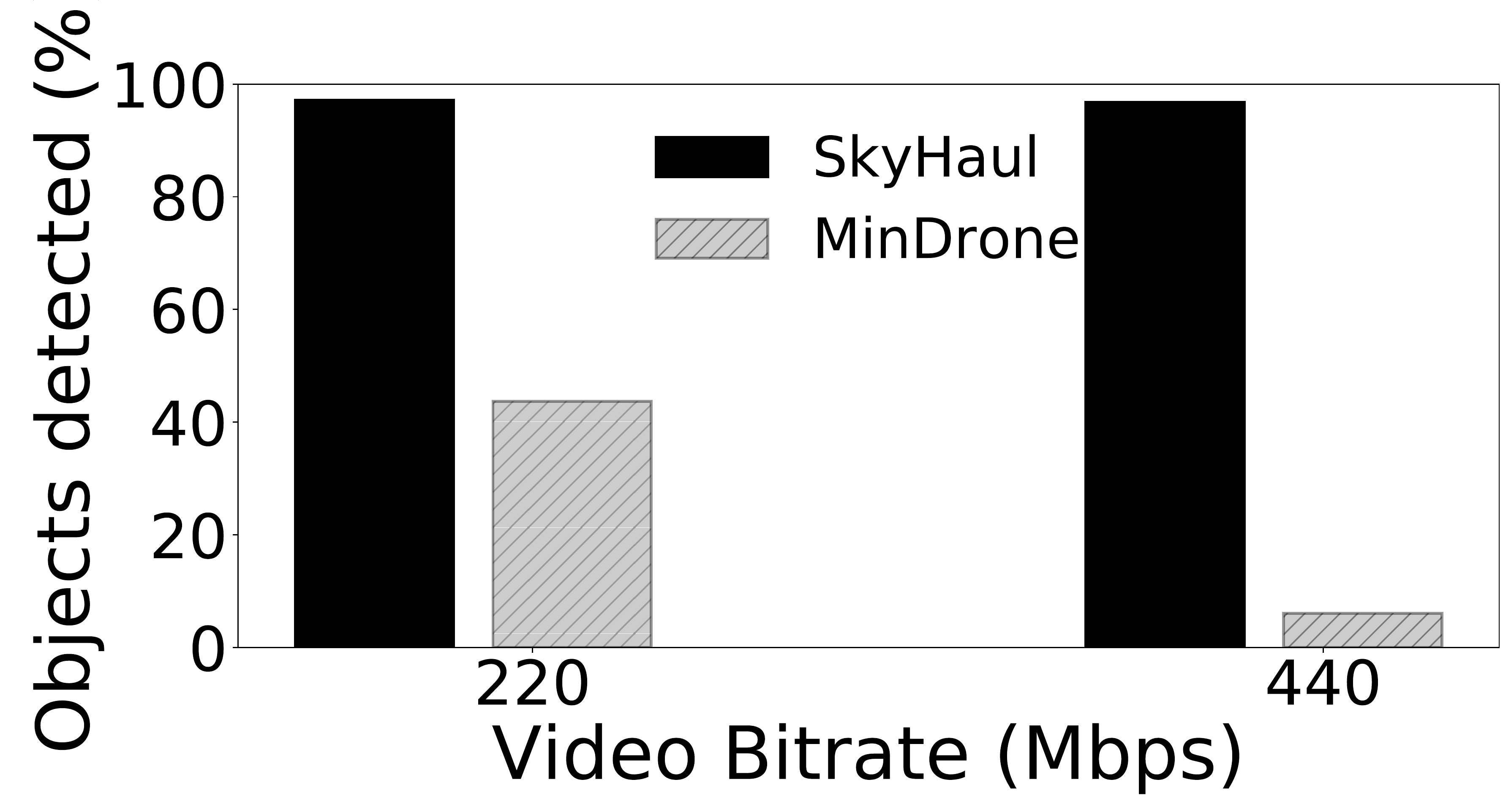}
        \caption {\small {Object detection in \system and \textsc{MinDrone}}  \label{fig:obj}}
    \end{subfigure}
    \vspace{-0.3cm}
    \caption {{Video processing: \textsc{MinDrone} vs. \system} \label{fig:eval_video}}
    \vspace{-0.5cm}
\end{figure}

\emph{\underline{\system accelerates video analytics:}} Fig.~\ref{fig:fps} shows
if we rely only on on-UAV compute resources, object detection proceeds at 0.2
frames-per-second (FPS). However, with \system's ability to maintain high
bandwidth links to the application UAVs, we achieve over 50$\times$ speed-up:
Object detection proceeds at up to 13FPS with a 440Mbps stream.

\emph{\underline{\system increases object detection accuracy:}} \system delivers
significantly fewer dropped frames to the cloud instance than \textsc{MinDrone}.
Fig.~\ref{fig:obj} shows that as a result, we achieve 97\% object detection
accuracy with \system (w.r.t. the lossless video recorded on the UAV) at all
bitrates while \textsc{MinDrone} can only achieve as low as 10\% detection accuracy 
(on a 440Mbps stream) due to lost frame data.

%% file: simulation.tex
The devastating forest fires in Paradise, CA~\cite{paradise_latimes,paradise_guardian},  
showed that unavailability of traditional 
communication infrastructure cell-towers can hamper first responders' rescue operations. 
We simulate deploying \system to provide backhaul connectivity to UAV based on-demand small-cells deployed during such emergencies.
To construct realistic topologies for application UAVs, we use the Paradise city map showing evacuation zones~\cite{paradise}. Assuming each zone
has at least one ground station at the center, we generate 140 topologies (14 zones, 10 per zone) each with 
5 (sparse deployment), 10 (moderate) and 20 (dense deployment) application UAVs within each zone. 
We consider 3 radios per relay UAV, and the ground station to have enough radio capacity to satisfy the net traffic 
demand of all application UAVs. Throughput requirement of each application UAV in each run is randomized to be between 100Mbps-1Gbps.

We compare the performance of \system against the following baseline algorithms:
(a)\textbf{\textsc{Steiner-MST}}: 
Finding Steiner trees that employ a minimal number of Steiner points (Relay UAVs) to connect the application UAVs with a bounded edge (transmission) length, is a NP-hard problem.  \textsc{Steiner-MST} provides a heuristic construction by finding the minimum spanning tree, connecting application UAVs to the GS UAV, then adding Relay UAVs to the paths computed based on the maximum communication length of each UAV. However, such Steiner solutions do not account for routing of traffic demands.   
(b)\textbf{\textsc{MaxCap}}: While \textsc{Steiner-MST} optimizes for connectivity with reduced UAV budget, \textsc{MaxCap} optimizes for capacity, to provide 100\% traffic satisfaction to each 
application UAV, without optimizing the number of relay UAVs employed. Here, the Relay UAV deployment is optimized for each application UAV in isolation, while accounting for other application UAVs that are sharing the same radio of the GS UAV.
(c)\textbf{\textsc{Air-Part}}: This is a subset of \system  with only the radial optimization (\S\ref{subsec:config}), i.e. no angular optimization across multiple sessions. 

\begin{figure}
    \centering
    \begin{subfigure}[b]{0.48\columnwidth}
      \includegraphics[width=\textwidth, height=2.9cm]{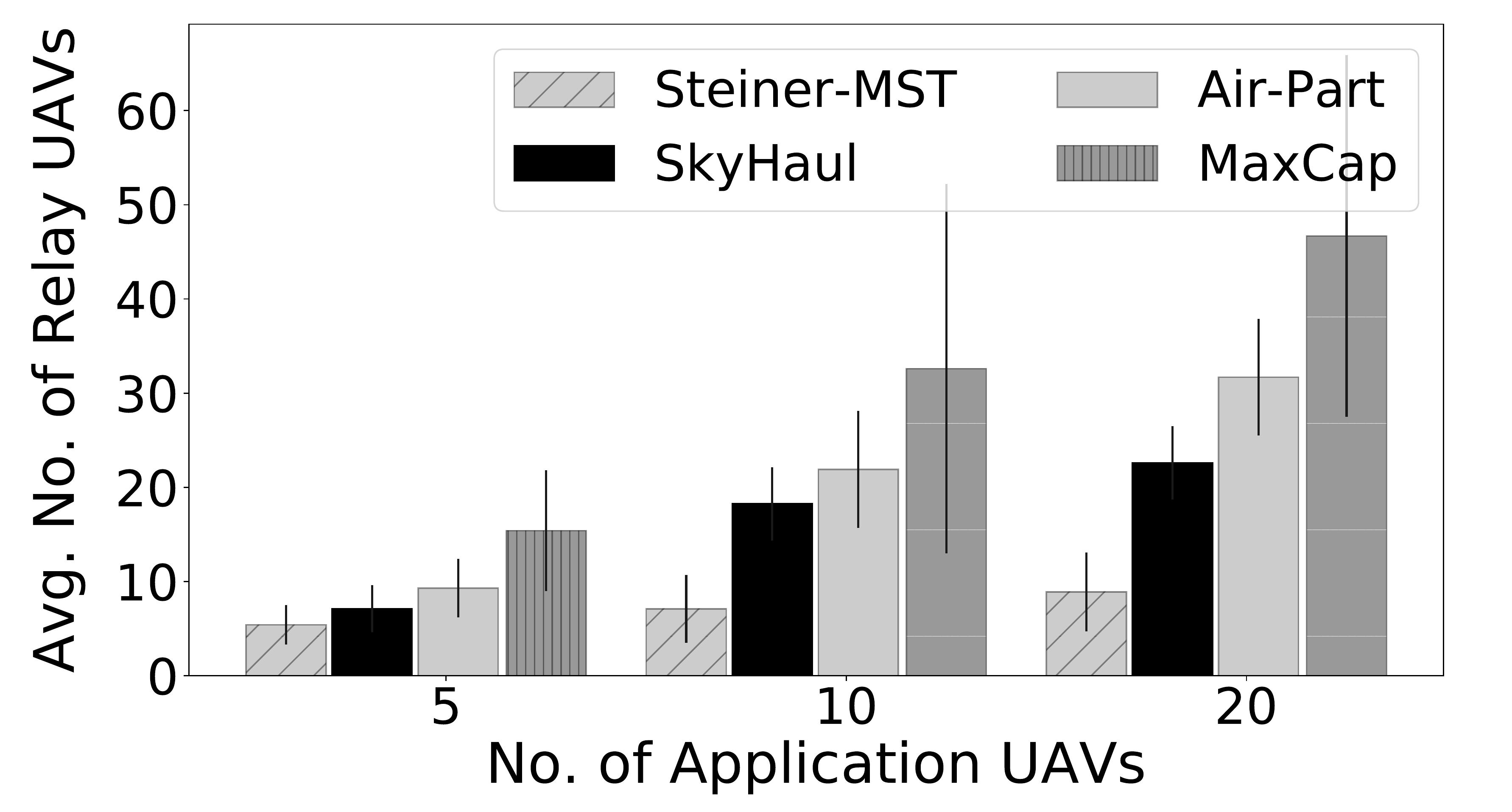}
    \caption {{No. of relay UAVs used \label{fig:no_drones}}}
    \end{subfigure}
    \begin{subfigure}[b]{0.48\columnwidth}
      \includegraphics[width=\textwidth, height=2.9cm]{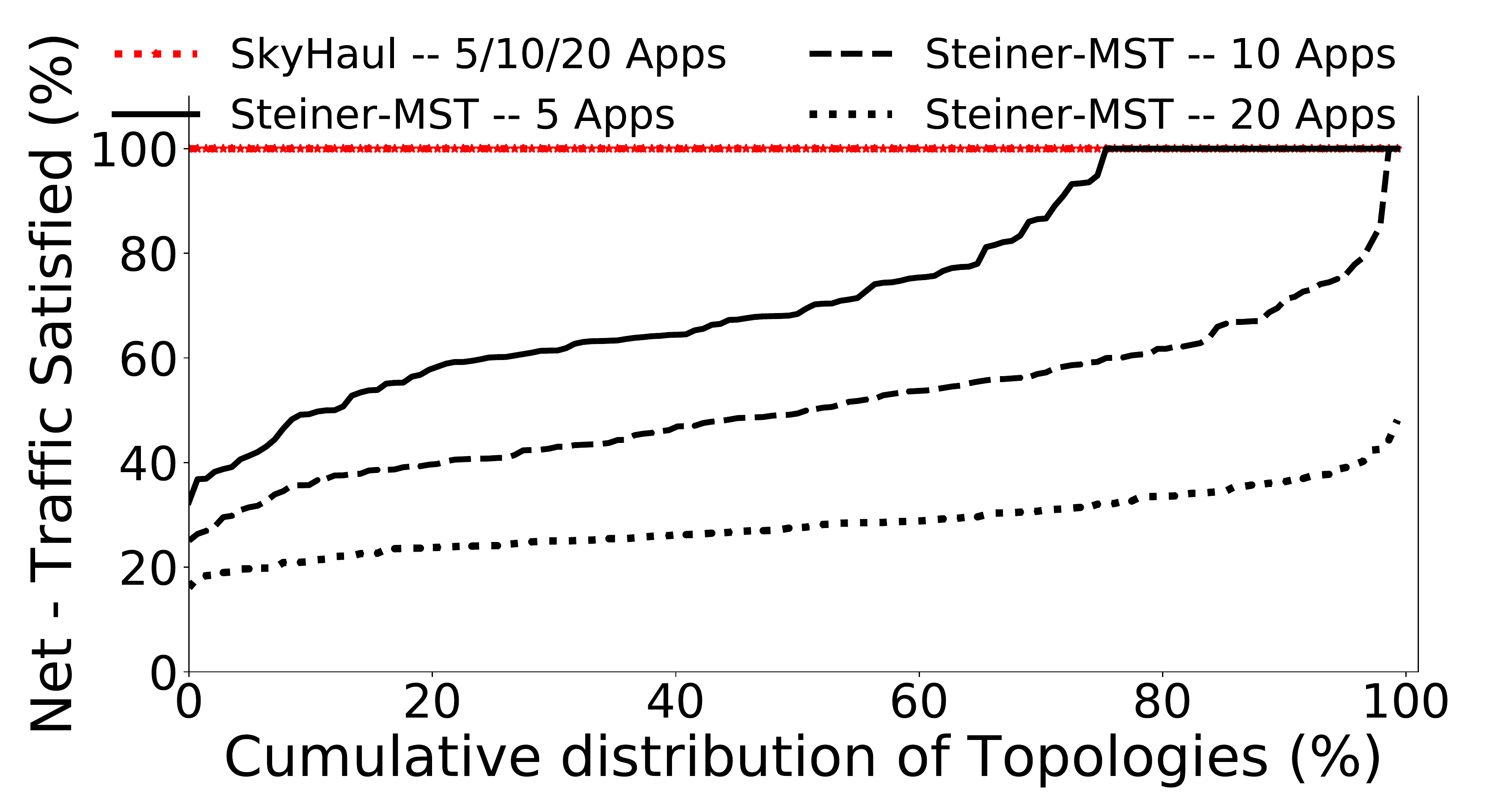}
    \caption {{Traffic satisfaction\label{fig:traf_sat}}}
    \end{subfigure}
    \vspace{-0.3cm}
    \caption{Simulation results for baselines and \system \label{fig:sim_res} }
    \vspace{-0.3cm}
\end{figure}

Fig.~\ref{fig:sim_res} shows the 
CDFs of traffic satisfaction (Net throughput / Net traffic demand) and average number of relay UAVs used by each algorithm.
While \textsc{Steiner-MST}
uses fewest UAVs, it is unable to satisfy the traffic requirement (Only 75\%, 45\% and 22\% traffic satisfied in median case
for 5, 10 and 20 application UAVs, respectively). On the other hand,
\textsc{MaxCap} ensures 100\% traffic satisfaction, but requires nearly four times the number of relay UAVs (for 20 Application UAVs) 
than \textsc{Steiner-MST}. 
In contrast, both \system and \textsc{Air-Part} strike a good balance between the number of relay UAVs needed and ensuring 100\% traffic satisfaction. While the radial optimization in 
\textsc{Air-Part} itself employs  30\% - 50\% fewer UAVs than the \textsc{MaxCap}, the angular contraction 
in \system optimizes deployment across sessions to reduce the 
UAVs required by 15\%-40\%, all while ensuring 100\% traffic satisfaction.

%% file: related.tex
{\bf 60GHz Link Performance:}
Many studies have focused on analyzing 60Ghz channel performance. 
Short-range 60Ghz link models for indoor locations~\cite{pfm:tap,saha:2017}
are not directly applicable for our reconfigurable environment. Outdoor 
60Ghz models~\cite{nit:2015,Haider1:2016}
developed using special hardwares like horn antennas and dedicated channel sounders
do not capture the mobility of UAVs and hence the
orientation of its mmWave radios on coarse time-scale network deployment decisions.

{\bf UAV based networks:} 
Apart from the popular high-altitude UAV network projects~\cite{loon,fb-aquila}, 
there has been recent interest in exploring solutions for low-altitude UAV-based cellular networks~\cite{att-cow,verizon}. 
\cite{mozaffari:2016, modares:2017, skyran, skycore}
use UAVs as stand-alone base stations, focusing 
on UAV placement for improving cellular coverage to ground users.
\system has a complementary focus in establishing a 
robust, mmWave-backhaul solution that can support the traffic demands of such cellular/WiFi access solutions. 
\cite{perez:2018, pinto:2017, sergio:2018} target multi-UAV networks, relying on numerical models and 
omni-directionality of the sub-6GHz RF technologies. This simplifies the UAV-deployment problem, while being 
unable to support high-bandwidth (mmWave) and the dynamics of applications.
\cite{kalantari:2017} investigates effects of different types of wireless backhauls offering various data rates on the number of ground users.

{\bf Routing algorithms:} Routing in multi-hop networks has been studied extensively~\cite{olsr,ett,etx},
with a limited focus on specific routing algorithms for Flying Adhoc Networks~\cite{sahingoz:2013}.
However, all these algorithms assume a pre-determined topology. This being not the case in \system, 
requires a joint optimization of both routing and topology.

%% file: conclusion2.tex
In this work, we presented the design and deployment of \system -- an autonomous, self-organizing, mmWave-meshed, multi-UAV network
that is capable of safely adapting itself to cater to highly dynamic applications and events. Through extensive real-world experiments, 
and large-scale simulations, we demonstrated the workings of \system.


{\bf \system for city-wide coverage:} In order to establish a city-wide backhaul network, the range of 
each mmWave radio needs to be longer than 360m supported in this work. However, radio products \cite{mikrotik-new}
that can extend range upto 1km are already available. \system can be easily scaled-up to cover cities using such long-range 
radios while reusing its existing algorithms. 




%% file: paper.bbl
\begin{thebibliography}{10}

\bibitem{us-ignite}
{US Ignite},
\newblock ``{Drone-Supported Emergency Management},''
  \url{https://www.us-ignite.org/apps/drone-supported-emergency-management/}.

\bibitem{utah-ignite}
{Utah Ignite},
\newblock ``{Utah Ignite -- Smart cities Smart Gigabit community},''
  \url{http://www.utahignite.com/}.

\bibitem{skyran}
Ayon Chakraborty, Eugene Chai, Karthikeyan Sundaresan, Amir Khojastepour, and
  Sampath Rangarajan,
\newblock ``Skyran: A self-organizing lte ran in the sky,''
\newblock in {\em Proceedings of the 14th International Conference on Emerging
  Networking EXperiments and Technologies}, New York, NY, USA, 2018, CoNEXT
  '18, pp. 280--292, ACM.

\bibitem{skycore}
Mehrdad Moradi, Karthikeyan Sundaresan, Eugene Chai, Sampath Rangarajan, and
  Z.~Morley Mao,
\newblock ``Skycore: Moving core to the edge for untethered and reliable
  uav-based lte networks,''
\newblock in {\em Proceedings of the 24th Annual International Conference on
  Mobile Computing and Networking}, New York, NY, USA, 2018, MobiCom '18, pp.
  35--49, ACM.

\bibitem{5g-VR}
{Qualcomm Technologies, Inc.},
\newblock ``{VR and AR pushing connectivity limits},''
  \url{https://www.qualcomm.com/media/documents/files/vr-and-ar-pushing-connectivity-limits.pdf}.

\bibitem{amazon-tower}
{Arthur Holland Michel},
\newblock ``{Amazon’s Drone Patents},''
  \url{https://dronecenter.bard.edu/files/2017/09/CSD-Amazons-Drone-Patents-1.pdf}.

\bibitem{loon}
{Alphabet Inc.,},
\newblock ``{Project Loon},'' \url{https://x.company/loon/.}, 2015.

\bibitem{foxsports}
{Fox},
\newblock ``Fox sports,''
  \url{https://www.foxsports.com/presspass/latest-news/2018/02/15/tethered-UAV-record-number-car-cameras-highlight-fox-sports-15th-daytona-500-broadcast}.

\bibitem{redcam}
{Red},
\newblock ``Red.com,'' \url{https://www.red.com/recording-time}.

\bibitem{olsr}
T.~Clausen and P.~Jacquet,
\newblock ``Optimized link state routing protocol (olsr),''
\newblock vol. 20, no. 3, 2003.

\bibitem{ett}
Richard Draves, Jitendra Padhye, and Brian Zill,
\newblock ``Routing in multi-radio, multi-hop wireless mesh networks,''
\newblock in {\em Proceedings of the 10th Annual International Conference on
  Mobile Computing and Networking}, New York, NY, USA, 2004, MobiCom '04, pp.
  114--128, ACM.

\bibitem{etx}
Douglas S.~J. De~Couto, Daniel Aguayo, John Bicket, and Robert Morris,
\newblock ``A high-throughput path metric for multi-hop wireless routing,''
\newblock in {\em Proceedings of the 9th Annual International Conference on
  Mobile Computing and Networking}, New York, NY, USA, 2003, MobiCom '03, pp.
  134--146, ACM.

\bibitem{microtik}
{Mikrotik},
\newblock ``Mikrotik,'' \url{https://mikrotik.com/product/wap_60g}.

\bibitem{aws}
{Amazon},
\newblock ``Amazon web service,'' \url{https://aws.amazon.com/}.

\bibitem{videodemo}
``Skyhaul demonstration flight,'' \url{https://tinyurl.com/twjcksz}.

\bibitem{Torkildson}
E.~{Torkildson}, H.~{Zhang}, and U.~{Madhow},
\newblock ``Channel modeling for millimeter wave mimo,''
\newblock in {\em 2010 Information Theory and Applications Workshop (ITA)},
  2010, pp. 1--8.

\bibitem{rappaport}
T.~S. {Rappaport}, Y.~{Xing}, G.~R. {MacCartney}, A.~F. {Molisch},
  E.~{Mellios}, and J.~{Zhang},
\newblock ``Overview of millimeter wave communications for fifth-generation
  (5g) wireless networks—with a focus on propagation models,''
\newblock {\em IEEE Transactions on Antennas and Propagation}, vol. 65, no. 12,
  pp. 6213--6230, 2017.

\bibitem{shahsavari}
S.~{Shahsavari}, M.~A.~A. {Khojastepour}, and E.~{Erkip},
\newblock ``Beam training optimization in millimeter-wave systems under
  beamwidth, modulation and coding constraints,''
\newblock in {\em 2019 IEEE 30th Annual International Symposium on Personal,
  Indoor and Mobile Radio Communications (PIMRC)}, 2019, pp. 1--7.

\bibitem{shahsavari2019}
Shahram Shahsavari and Elza Erkip,
\newblock ``Robust beam tracking and data communication in millimeter wave
  mobile networks,''
\newblock 2019.

\bibitem{haider:2016}
Muhammad~Kumail Haider and Edward~W. Knightly,
\newblock ``Mobility resilience and overhead constrained adaptation in
  directional 60 ghz wlans: Protocol design and system implementation,''
\newblock in {\em Proceedings of the 17th ACM International Symposium on Mobile
  Ad Hoc Networking and Computing}, New York, NY, USA, 2016, MobiHoc ’16, p.
  61–70, Association for Computing Machinery.

\bibitem{uav4}
{Digital Trends},
\newblock ``{7 Drones that can stay airborne for hours — and the tech that
  makes it possible},''
  \url{https://www.digitaltrends.com/cool-tech/drones-with-super-long-flight-times/}.

\bibitem{chen:2000}
Donghui chen, Ding-Zhu Du, xiao-dong Hu, guo-hui Lin, Lusheng Wang, and
  guoliang Xue,
\newblock ``Approximations for steiner trees with minimum number of steiner
  points,''
\newblock {\em Journal of Global Optimization}, vol. 18, no. 1, pp. 17--33, Sep
  2000.

\bibitem{uav1}
{Skyfront},
\newblock ``{Long endurance Drone},'' \url{https://skyfront.com/}.

\bibitem{uav2}
{Quaternium},
\newblock ``{Hybrid Drone},'' \url{http://www.quaternium.com/}.

\bibitem{uav3}
{UAV Factory},
\newblock ``{Long endurance UAS, UAVs and subsystems},''
  \url{https://www.unmannedsystemstechnology.com/company/uav-factory/}.

\bibitem{yolo3}
Joseph Redmon, Santosh~Kumar Divvala, Ross~B. Girshick, and Ali Farhadi,
\newblock ``You only look once: Unified, real-time object detection,''
\newblock {\em CoRR}, vol. abs/1506.02640, 2015.

\bibitem{ssim}
and A.~C.~{Bovik}, H.~R. {Sheikh}, and E.~P. {Simoncelli},
\newblock ``Image quality assessment: from error visibility to structural
  similarity,''
\newblock {\em IEEE Transactions on Image Processing}, vol. 13, no. 4, pp.
  600--612, April 2004.

\bibitem{paradise_latimes}
{Los Angeles Times},
\newblock ``{Paradise City Forest Fires},''
  \url{https://www.latimes.com/local/california/la-me-camp-fire-deathtrap-20181230-story.html},
  2018.

\bibitem{paradise_guardian}
{The Guardian},
\newblock ``{Paradise City Forest Fires},''
  \url{https://www.theguardian.com/environment/2018/dec/20/last-day-in-paradise-california-deadliest-fire-untold-story-survivors},
  2018.

\bibitem{paradise}
{City of Paradise},
\newblock ``Paradise, ca,''
  \url{https://www.townofparadise.com/index.php/17-news-events/248-evacuation-zones}.

\bibitem{pfm:tap}
P.~F.~M. Smulders,
\newblock ``Statistical characterization of 60-ghz indoor radio channels,''
\newblock {\em Transactions on Antennas and Propagation}, vol. 57, no. 10, pp.
  2820--2829, Oct. 2009.

\bibitem{saha:2017}
S.~K. {Saha}, T.~{Siddiqui}, D.~{Koutsonikolas}, A.~{Loch}, J.~{Widmer}, and
  R.~{Sridhar},
\newblock ``A detailed look into power consumption of commodity 60 ghz
  devices,''
\newblock in {\em 2017 IEEE 18th International Symposium on A World of
  Wireless, Mobile and Multimedia Networks (WoWMoM)}, June 2017, pp. 1--10.

\bibitem{nit:2015}
Thomas Nitsche, Guillermo Bielsa, Irene Tejado, Adrian Loch, and Joerg Widmer,
\newblock ``Boon and bane of 60 ghz networks: practical insights into
  beamforming, interference, and frame level operation,''
\newblock in {\em ACM CoNEXT 2015}, 12 2015, pp. 1--13.

\bibitem{Haider1:2016}
Muhammad~Kumail Haider and Edward~W. Knightly,
\newblock ``Mobility resilience and overhead constrained adaptation in
  directional 60 ghz wlans: Protocol design and system implementation,''
\newblock in {\em Proceedings of the 17th ACM International Symposium on Mobile
  Ad Hoc Networking and Computing}, New York, NY, USA, 2016, MobiHoc '16, pp.
  61--70, ACM.

\bibitem{fb-aquila}
{Facebook},
\newblock ``{Facebook Project Aquilla},''
  \url{https://code.facebook.com/posts/348442828901047/aquila-what-s-next-for-high-altitude-connectivity-/}.

\bibitem{att-cow}
{ATT networks},
\newblock ``{ATT Cell On Wings},''
  \url{http://about.att.com/innovationblog/cows_fly.}

\bibitem{verizon}
{Verizon networks},
\newblock ``{Verizon cell on wings},'' \url{”
  https://newatlas.com/verizon-drones-internet-trials/45818/.}

\bibitem{mozaffari:2016}
M.~{Mozaffari}, W.~{Saad}, M.~{Bennis}, and M.~{Debbah},
\newblock ``Efficient deployment of multiple unmanned aerial vehicles for
  optimal wireless coverage,''
\newblock {\em IEEE Communications Letters}, vol. 20, no. 8, pp. 1647--1650,
  Aug 2016.

\bibitem{modares:2017}
J.~{Modares}, F.~{Ghanei}, N.~{Mastronarde}, and K.~{Dantu},
\newblock ``Ub-anc planner: Energy efficient coverage path planning with
  multiple drones,''
\newblock in {\em 2017 IEEE International Conference on Robotics and Automation
  (ICRA)}, May 2017, pp. 6182--6189.

\bibitem{perez:2018}
A~Guillen-Perez and M.D Cano,
\newblock ``Flying ad hoc networks: A new domain for network communications,''
\newblock 2018, Sensors, pp. 958--963.

\bibitem{pinto:2017}
L.~R. {Pinto}, A.~{Moreira}, L.~{Almeida}, and A.~{Rowe},
\newblock ``Characterizing multihop aerial networks of cots multirotors,''
\newblock {\em IEEE Transactions on Industrial Informatics}, vol. 13, no. 2,
  pp. 898--906, April 2017.

\bibitem{sergio:2018}
S{\'e}rgio Sabino and Ant\'{o}nio Grilo,
\newblock ``Topology control of unmanned aerial vehicle (uav) mesh networks: A
  multi-objective evolutionary algorithm approach,''
\newblock in {\em Proceedings of the 4th ACM Workshop on Micro Aerial Vehicle
  Networks, Systems, and Applications}, New York, NY, USA, 2018, DroNet'18, pp.
  45--50, ACM.

\bibitem{kalantari:2017}
E.~{Kalantari}, M.~Z. {Shakir}, H.~{Yanikomeroglu}, and A.~{Yongacoglu},
\newblock ``Backhaul-aware robust 3d drone placement in 5g+ wireless
  networks,''
\newblock in {\em 2017 IEEE International Conference on Communications
  Workshops (ICC Workshops)}, May 2017, pp. 109--114.

\bibitem{sahingoz:2013}
Ozgur~Koray Sahingoz,
\newblock ``Networking models in flying ad-hoc networks ({FANETs}): Concepts
  and challenges,''
\newblock {\em Journal of Intelligent {\&} Robotic Systems}, vol. 74, no. 1-2,
  pp. 513--527, oct 2013.

\bibitem{mikrotik-new}
{Mikrotik},
\newblock ``Long range 60ghz radio,''
  \url{https://mikrotik.com/product/lhg_lite60}.

\end{thebibliography}
